\documentclass[twocolumn]{aastex62}

\usepackage{soul}
\graphicspath{{./}{figures/}}

\submitjournal{ApJ}

\shorttitle{Peter Pan Disks}
\shortauthors{Silverberg et al.}
\begin{document}

\title{Peter Pan Disks: Long-lived Accretion Disks Around Young M Stars}

\author{Steven M. Silverberg}\altaffiliation{Visiting astronomer, Cerro Tololo Inter-American Observatory, \\National Optical Astronomy Observatory, which is operated by \\the Association of Universities for Research in Astronomy \\(AURA) under a cooperative agreement with the National \\Science Foundation.}
%\affiliation{Homer L. Dodge Department of Physics and Astronomy, University of Oklahoma, 440 W. Brooks Street, Norman, OK 73019, USA}
%\affiliation{NASA Goddard Space Flight Center, Exoplanets and Stellar Astrophysics Laboratory, Code 667, Greenbelt, MD 20771}
\affiliation{MIT Kavli Institute for Astrophysics and Space Research, 77 Massachusetts Avenue, Cambridge, MA, 02139, USA}

\author{John P. Wisniewski}
\affiliation{Homer L. Dodge Department of Physics and Astronomy, University of Oklahoma, 440 W. Brooks Street, Norman, OK 73019, USA}

\author{Marc J. Kuchner}
\affiliation{NASA Goddard Space Flight Center, Exoplanets and Stellar Astrophysics Laboratory, Code 667, Greenbelt, MD 20771}

\author{Kellen D. Lawson}
\affiliation{Homer L. Dodge Department of Physics and Astronomy, University of Oklahoma, 440 W. Brooks Street, Norman, OK 73019, USA}

\author{Alissa S. Bans}
\affiliation{Department of Physics, Emory University, 201 Dowman Drive, Atlanta, GA 30322, USA}

\author{John H. Debes}
\affiliation{Space Telescope Science Institute, 3700 San Martin Dr., Baltimore, MD 21218, USA}

\author{Joseph R. Biggs}
\affiliation{Disk Detective Citizen Scientist}

\author{Milton K.D. Bosch}
\affiliation{Disk Detective Citizen Scientist}

\author{Katharina Doll}
\affiliation{Disk Detective Citizen Scientist}

\author{Hugo A. Durantini Luca}
\affiliation{Disk Detective Citizen Scientist}
\affiliation{IATE-OAC, Universidad Nacional de C\'ordoba-CONICET. Laprida 854, X5000 BGR, C\'ordoba, Argentina}

\author{Alexandru Enachioaie}
\affiliation{Disk Detective Citizen Scientist}

\author{Joshua Hamilton}
\affiliation{Disk Detective Citizen Scientist}

\author{Jonathan Holden}
\affiliation{Disk Detective Citizen Scientist}

\author{Michiharu Hyogo}
\affiliation{Disk Detective Citizen Scientist}
\affiliation{Meisei University, 2-1-1 Hodokubo, Hino, Tokyo 191-0042, Japan}

\collaboration{The Disk Detective Collaboration}

\begin{abstract}
%{\bf These first two sentences aren't very well connected; one is about debris disks, one is about accretion disks.  It would take a few more sentences to explain the connection.  I propose that instead we delete the first sentence (about debris disks). -MJK}
%Despite the prevalence of detected M dwarf exoplanets and M-type primordial disks, few M dwarf debris disks have been detected.
WISEA J080822.18-644357.3, an M star in the Carina association, exhibits extreme infrared excess and accretion activity at an age greater than the expected accretion disk lifetime. We consider J0808 as the prototypical example of a class of M star accretion disks at ages $\gtrsim 20$ Myr, which we call ``Peter Pan'' disks, since they apparently refuse to grow up. We present four new Peter Pan disk candidates identified via the Disk Detective citizen science project, coupled with \textit{Gaia} astrometry. We find that WISEA J044634.16-262756.1 and WISEA J094900.65-713803.1 both exhibit significant infrared excess after accounting for nearby stars within the 2MASS beams. The J0446 system has $>95\%$ likelihood of Columba membership. The J0949 system shows $>95\%$ likelihood of Carina membership. We present new GMOS optical spectra of all four objects, showing possible accretion signatures on all four stars. We present ground-based and \textit{TESS} lightcurves of J0808 and 2MASS J0501-4337, including a large flare and aperiodic dipping activity on J0808, and strong periodicity on J0501. We find Pa$\beta$ and Br$\gamma$ emission indicating ongoing accretion in near-IR spectroscopy of J0808. Using observed characteristics of these systems, we discuss mechanisms that lead to accretion disks at ages $\gtrsim20$ Myr, and find that these objects most plausibly represent long-lived CO-poor primordial disks, or ``hybrid'' disks, exhibiting both debris- and primordial-disk features. The question remains: why have gas-rich disks persisted so long around these particular stars?  %{\bf NOTE: Consider using "archetype" instead of "prototype" throughout the paper. -MJK}
%{\bf I punched up the end of this. -MJK}
\end{abstract}

%% Keywords should appear after the \end{abstract} command. 
%% See the online documentation for the full list of available subject
%% keywords and the rules for their use.

%% From the front matter, we move on to the body of the paper.
%% Sections are demarcated by \section and \subsection, respectively.
%% Observe the use of the LaTeX \label
%% command after the \subsection to give a symbolic KEY to the
%% subsection for cross-referencing in a \ref command.
%% You can use LaTeX's \ref and \label commands to keep track of
%% cross-references to sections, equations, tables, and figures.
%% That way, if you change the order of any elements, LaTeX will
%% automatically renumber them.
%%
%% We recommend that authors also use the natbib \citep
%% and \citet commands to identify citations.  The citations are
%% tied to the reference list via symbolic KEYs. The KEY corresponds
%% to the KEY in the \bibitem in the reference list below. 

\section{Introduction}

Primordial disks around solar-mass stars have been shown to dissipate rapidly over time in \textit{Spitzer} disk surveys. The From Molecular Cores to Planet-Forming Disks (c2d) project observed 188 classical and weak T Tauri stars (CTTS; WTTS) spanning spectral types G5-M5 in the Taurus, Lupus, Ophiuchus, and Chamaeleon star-forming regions \citep{2006ApJ...645.1283P} and found that no studied weak-lined T Tauri stars (WTTS) exhibited a disk beyond an age of 10 Myr \citep{2010ApJ...724..835W}. The Formation and Evolution of Planetary Systems (FEPS) project surveyed $\sim 328$ approximately solar-mass ($0.7-2.2 M_{\odot}$) stars from ages 3 Myr to 3 Gyr and found that only $\sim 12\%$ of stars surveyed younger than 10 Myr exhibited a primordial disk, and only $\sim 2\%$ of stars surveyed between 10 and 30 Myr exhibited such a disk \citep{2009ApJS..181..197C}. 

Studies of large young clusters and associations have also shown that disks around high-mass star disks dissipate more quickly than those around solar-mass stars \citep{2006ApJ...651L..49C}. \citet{2016MNRAS.461..794P} performed a census of disks around K stars in the Scorpius-Centaurus complex and determined disk fractions for each component of Sco-Cen, finding full disk fractions of $9.0_{-2.2}^{+4.0}\%$  for the $10 \pm 3$ Myr Upper Scorpius association, $5.1_{-1.2}^{+2.4}\%$ for the $16 \pm 2$ Myr Upper Centaurus-Lupus (UCL) association, and $3.4_{-1.0}^{+2.5}\%$ for the $15 \pm 3$ Myr Lower Centaurus-Crux (LCC) association. Using these fractions, they inferred that the characteristic ($e$-folding) timescale for primordial disk lifetimes around K-type stars was 4.7 Myr---that is, only 1/$e$ of the primordial disks around K-type stars in this association have not dissipated in a 4.7-Myr span. 

Given that K-type stars have higher primordial disk fractions than solar-type stars in the same associations, one might expect that M dwarf primordial disks would have an even longer lifetime. However, there is to date only one detailed study of M dwarf disks in only one association, the disk census of Upper Scorpius by \citet{2018AJ....156...75E}. This study did find that the frequency of M dwarf primordial disks in Upper Sco is higher than the frequency of primordial disks around all earlier-type stars in the same association, which could indicate a longer disk lifetime; however, further studies of additional associations at different ages (e.g. a similar census of LCC and UCL) are needed to confirm that this is the case.

In contrast to the prevalence of M star and brown dwarf primordial disks in young associations, M dwarf {\it debris disks} are detected much less frequently than their counterparts around higher-mass stars. Debris disks are clouds of primarily rock and dust, thought to be a later stage of disk evolution than primordial disks \citep[e.g.][]{2018ARA&A..56..541H}. 
The occurrence rate around field M dwarfs is $<1.4\%$, based on surveys with the \textit{Spitzer Space Telescope} and the \textit{WISE} All-Sky Catalog \citep{Plavchan2005,Plavchan2009,2012A&A...548A.105A}. These systems are more commonly detected in younger systems; \citet{2008ApJ...687.1107F} found significant 24$\mu$m excess around 4.3\% of M dwarfs in the 30-40 Myr cluster NGC 2547, while \citet{2017MNRAS.469..579B} found significant 22 $\mu$m excess around $13\% (\pm5\%)$ of $<30$-Myr M dwarfs. These detection rates are still significantly lower than the $\sim 70\%$ detection rates of their earlier-type A-star counterparts \citep{2010MNRAS.409L..44G}. While some argue that this could be due to the sensitivity of WISE, with dust material harder to detect due to the low intrinsic luminosity of the host stars \citep[e.g.][]{2013MNRAS.432.2562H}, the low detection rate of debris disks around older M dwarfs could also indicate rapid clearing by stellar wind \citep[e.g.][]{2008ARA&A..46..339W} or stellar activity (e.g. AU Mic; Grady et al. 2019, in prep). 

The Disk Detective citizen science project \citep{Kuchner2016}, a joint program of NASA and the Zooniverse Project \citep{2008MNRAS.389.1179L}, identifies circumstellar disk candidates in data from NASA's WISE mission, via visual inspection of candidates in the AllWISE catalog. \citet{2016ApJ...830L..28S} identified WISEA J080822.18-644357.3 (J0808), a disk candidate discovered by Disk Detective citizen scientists, as a likely member of the 45-Myr Carina association \citep{2015MNRAS.454..593B} based on its kinematics. J0808 exhibits a large fractional infrared luminosity ($L_{\mathrm{ir}}/L_{\star} > 0.01$), surprising given its age, as such a high $L_{\mathrm{ir}}/L_{\star}$ is typically only seen in primordial disks. \citet{2016ApJ...830L..28S} initially characterized this as one of the oldest dM-type debris disk systems. \citet{2018MNRAS.476.3290M} confirmed membership of J0808 in Carina based on its radial velocity and lithium absorption detected in optical spectroscopy, and noted broad, variable H$\alpha$ emission indicative of active accretion, suggesting that the system is instead a gas-rich primordial disk. They also identified correlated variability in the W1 and W2 single-epoch photometry of the object consistent with a variable hot disk component, and listed three other systems from the literature that exhibited emission indicative of accretion \citep[2MASS J0041353-562112, J02265658-5327032, and 2MASS J05010082-4337102, hereafter J0041, J0226, and J0501;][]{2009ApJ...702L.119R,2016ApJ...832...50B} in the Tucana-Horologium (THA), Columba (COL), and Carina (CAR) associations, which formerly made up the Great Austral Young Association (GAYA) complex \citep{2008hsf2.book..757T}. While these groups have all been thought to have ages $\sim 45$ Myr \citep{2015MNRAS.454..593B}, recent work suggests that Carina may have an age closer to that of the $~24$-Myr Beta Pictoris Moving Group \citep{2019AJ....157..234S}.
\citet{2019ApJ...872...92F} recently published ALMA observations of J0808, in which they detected continuum dust emission consistent with a third disk component (in addition to those identified in the WISE data), but did not detect cold CO gas.

%{\bf This first sentence is a run-on and needs some more thought.  It's an important sentence because it contains your definition of Peter Pan Disks. Maybe delete "based on membership in a $\sim45$ Myr moving group"  since I think we would be happy with any moving group older than 35 Myr.   Also we need to be quantitative about "substantial" IR excess.  Should we place the limit at LIR/Lstar = 1 percent since that's often used as the dividing line between debris and primordial disks?   Why is the age limit 35 Myr, by the way? If you don't have a reason for the number, leave yourself some wiggle room in the definition. Also, let's do something to set our definition apart from the rest of the text. Maybe put the requirements (older than 35 Myr, high LIR/Lstar, etc.) into bullet points? Alternatively, replace this with a statement saying that you will define this class of disks later (in what section?)- MJK}
In this paper, we discuss the characteristics of the known examples of a class of disk we name ``Peter Pan'' disks: disks around low-mass stars and brown dwarfs that exhibit characteristics of a gas-rich disk at unexpectedly high ages---that is, accretion disks that seem to ``never grow up'' \citep{Barrie1904}. In Section \ref{sec:Gaia}, we present identifications of four new examples of this phenomenon in two systems, based on astrometry and kinematic information from the second data release of \textit{Gaia}. In Section \ref{sec:new_disks_observations}, we summarize follow-up spectroscopy of these systems with Gemini/GMOS. In Section \ref{sec:new_disks_analysis}, we combine the observed photometry, \textit{Gaia} astrometry, and GMOS spectra of these targets to assess their ages and accretion properties. In Section \ref{sec:observations}, we summarize our observations of the previously-known Peter Pan disks, including ground- and space-based high-cadence photometry and near-infrared spectroscopy. In Section \ref{sec:lightcurves}, we present high-cadence optical photometry of J0808, which exhibits variability indicative of accretion as well as a high-amplitude classical flare, and J0501, which indicates strong periodicity and a classical flare. In Section \ref{sec:spectra}, we present near-IR spectroscopy of J0808, which identifies the short-wavelength end of the W1/2 excess identified by \citet{2018MNRAS.476.3290M} and shows variable accretion. In Section \ref{sec:mechanisms}, we examine a possible definition of the class of Peter Pan disks, consider what other known systems might belong in this class, and discuss potential formation mechanisms for these systems. We summarize our findings in Section \ref{sec:conclusion}.

\section{Two New Peter Pan Disk Systems}
\label{sec:Gaia}

As part of our ongoing work to build a publicly-accessible database for Disk Detective through the Mikulski Archive for Space Telescopes at the Space Telescope Science Institute\footnote[1]{Accessible through the portal at \url{https://mast.stsci.edu}.}, we cross-matched the \textit{Gaia} DR2 catalog with the Disk Detective input catalog. Rather than exclusively using the matches provided in DR2, we projected objects in DR2 back to their expected WISE positions using the proper motion data in DR2, then found the nearest WISE source to that position. Our cross-match did not explicitly take into account the position of nearby objects in the \textit{Gaia} data. We ran the results of this crossmatch through the BANYAN $\Sigma$ \citep{2018ApJ...856...23G} tool to identify potential new moving group members, specifically with an eye toward identifying new Peter Pan disks. This produced two late-type systems with a high probability of membership in a $\sim45$-Myr moving group, based on their parallax and proper motion from \textit{Gaia}. WISEA J044634.16-262756.1 has a 96.9\% likelihood of membership in the $\sim42$-Myr Columba association (COL). WISEA J094900.65-713803.1 has a 99.2\% likelihood of membership in the Carina association. Note that these initial probabilities do not reflect radial velocity measurements, which we discuss below.

\begin{figure*}
\plottwo{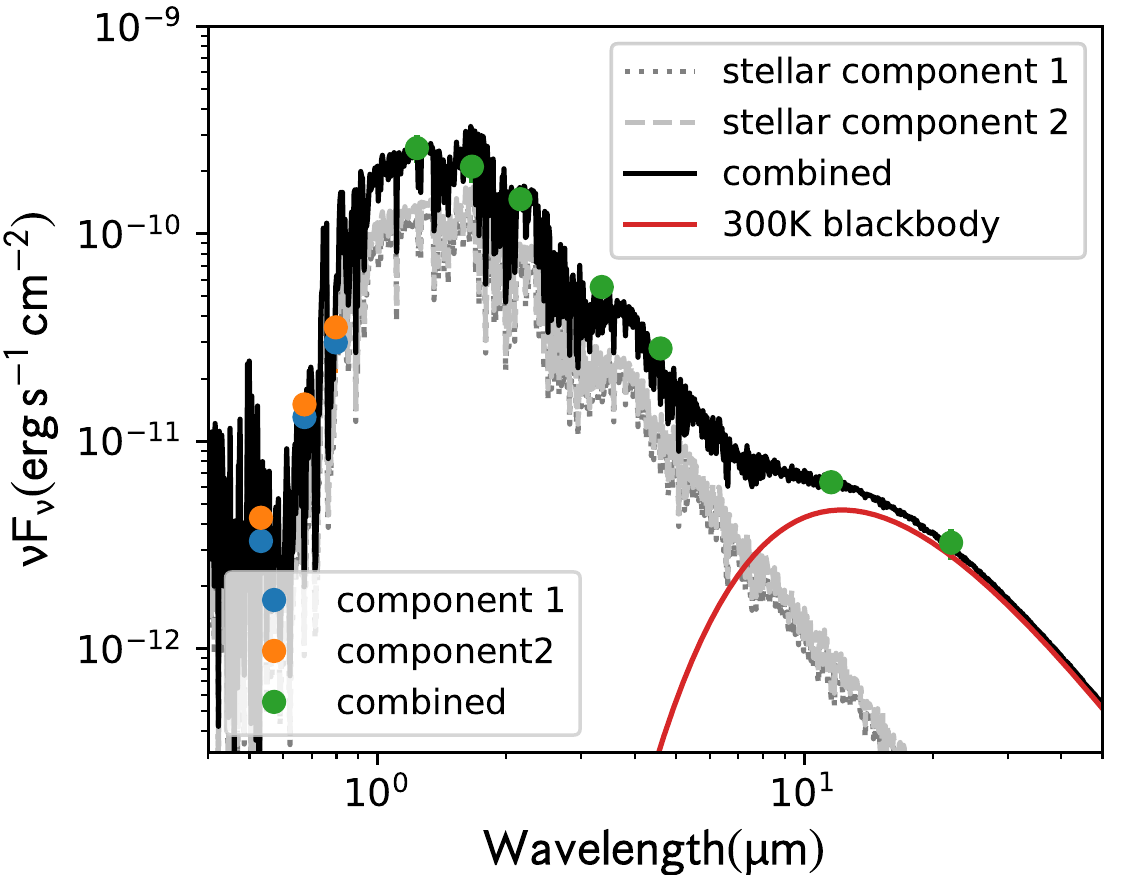}{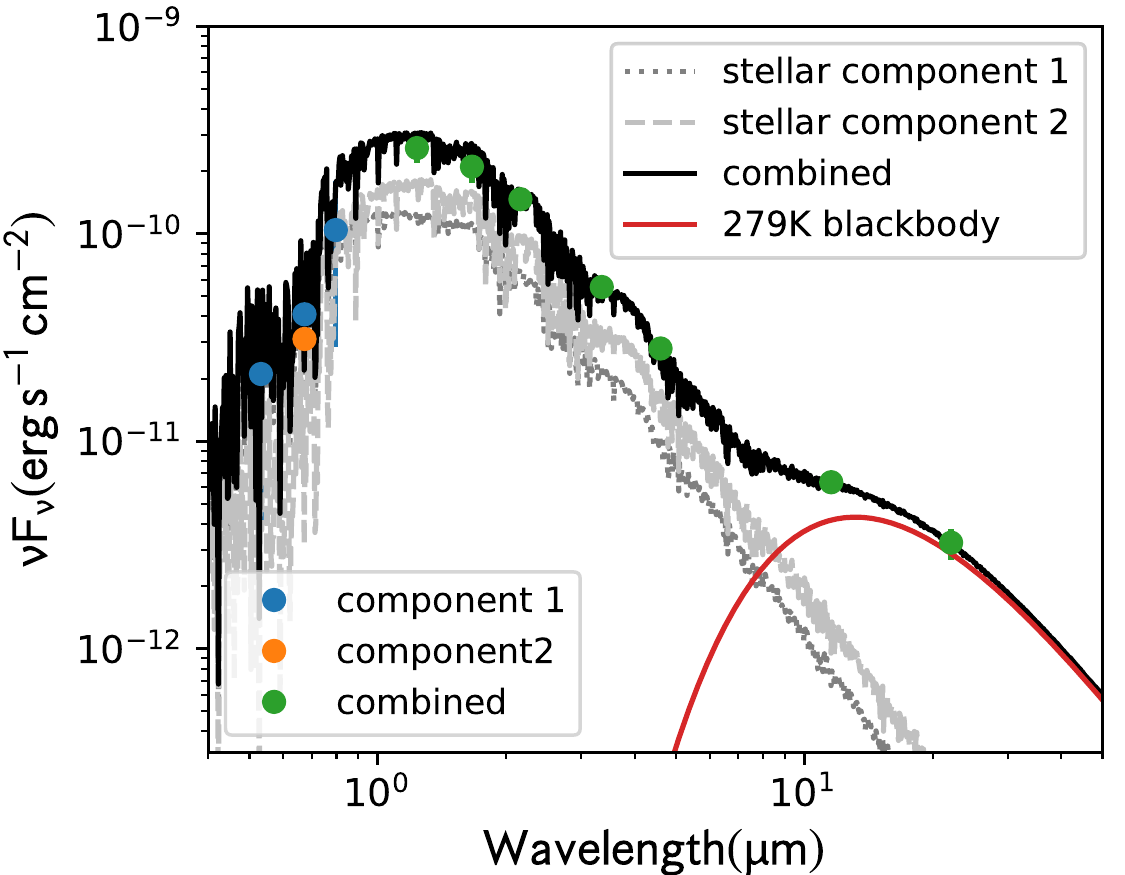}
\caption{\textit{Left}: SED for the J0446 system, compared to models for the spectra of both stars and blackbody dust emission that have been fit to the photometry. \textit{Right}: SED and models for the J0949 system. Characteristics of the systems are listed in Table \ref{table:new_disks}.}
\label{fig:new_SEDS}
\end{figure*}

Each of these systems is an apparent visual double, based on data from Pan-STARRS DR1 and \textit{Gaia} DR2. The two stars in J0446 have angular separation $2''.3$, while the two stars in J0949 have angular separation $1''.5$; both of these separations are within the beam size of both 2MASS and WISE. To characterize these systems as a whole, we fit stellar models to the observed photometry for the systems, using the \textit{Gaia} photometry for each component and the 2MASS/WISE photometry that blends both components. Three of the targets have \textit{Gaia} photometry in all three bands (G, Bp, Rp), while J0949B only has G-band photometry. These initial fits to the spectral energy distributions (SEDs) of the systems indicated that both J0446 and J0949 exhibited significant excess at W3 and W4 after accounting for both stellar components. Additionally, BANYAN $\Sigma$ indicated that the two nearby targets were also possible moving group members. SED fits of the systems are shown in in Figure \ref{fig:new_SEDS}.

\section{Follow-up Spectroscopy of New Peter Pan Disk Systems}
\label{sec:new_disks_observations}
\begin{figure*}
\includegraphics[width=0.97\textwidth]{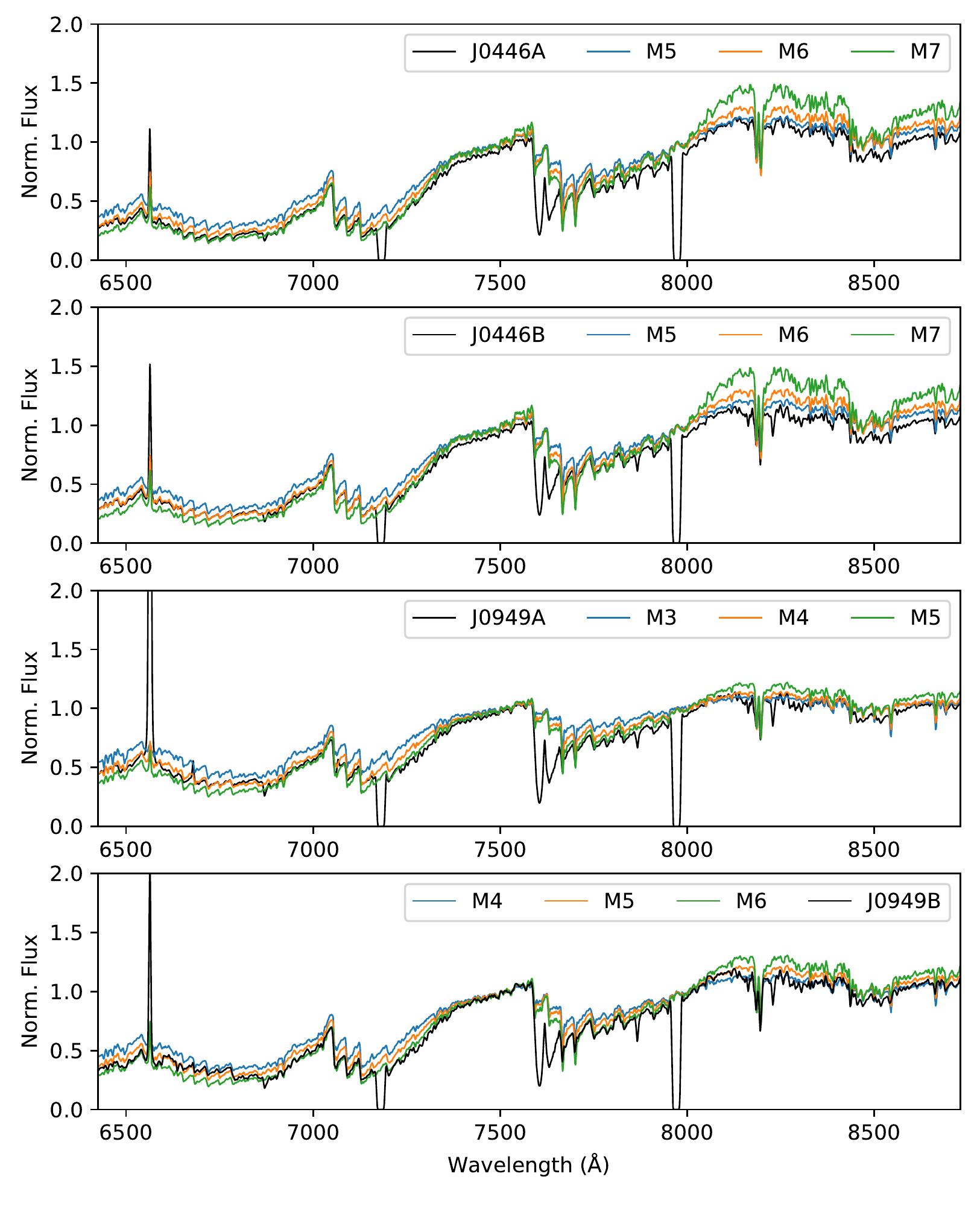}
\caption{Spectra of the four new potential Peter Pan disks, plotted against PyHammer empirical spectral templates. Observed and template spectra are normalized at 7500\AA. The dips in the observed spectra are due to the GMOS chip gaps. Note the strong H$\alpha$ emission in all four spectra, and the He I $\lambda6678$ emission in J0949A.}
\label{fig:gmos_spectra}
\end{figure*}

We obtained spectra of these two new candidate Peter Pan disk systems during 2018 November with the Gemini Multi-Object Spectrograph (GMOS-S) at the Gemini-South observatory through Program GS-2018B-FT-106 in queue mode, to check moving group memberships and further characterize the systems. We observed both objects in each apparent visual double. We acquired three spectra of each target in sequence to mitigate cosmic rays, using the $0''.5$ longslit with the R831 grating centered at 7570 \AA\ to achieve resolving power $\sim 4400$, with wavelength coverage of 6394-8736 \AA. We used the GG455 filter to block second-order light. Quartz/halogen flats were taken after each set of science observations, and master arcs for the project were taken using Cu-Ar lamps on 2018 November 8. We used the flux standard LTT 1020, observed on 2018 October 3, to calibrate our result spectra. Exposure times ranged from 12-100 seconds, based on expected brightness, yielding spectra with SNR 14-25 at H$\alpha$. Spectra were reduced using version 2.16 of the Gemini package in PyRAF, using standard techniques. Table \ref{table:new_disks} lists the four components as J0446A, J0446B, J0949A, and J0949B, along with separate equatorial positions for each target, astrometric information, \textit{Gaia} photometry, and data derived from the spectra. The spectra are plotted in Figure \ref{fig:gmos_spectra}.

\begin{deluxetable*}{lccccc}
\tablewidth{0pt}
\tabletypesize{\footnotesize}
\tablecaption{New Peter Pan Disk Candidates From \textit{Gaia} DR2 and BANYAN $\Sigma$ \label{table:new_disks}}
\tablehead{\colhead{Designation} & \colhead{J0446A} & \colhead{J0446B} & \colhead{J0949A} & \colhead{J0949B} & \colhead{References}}
\startdata
R.A. (h:m:s) & 04:46:34.105362 & 04:46:34.249381 & 09:49:00.752604 & 09:49:00.441275 & 1\\
DEC (d:m:s) & -26:27:56.83936 & -26:27:55.57007 & -71:38:02.94748 & -71:38:03.15884 & 1\\
Separation (pc) & \multicolumn{2}{c}{$0.347 \pm 0.341$} & \multicolumn{2}{c}{$1.100 \pm 0.474$} & 2 \\
Spectral Type & M6 & M6 & M4 & M5 & 2\\
Group & COL & COL & CAR & CAR & 3\\
EW[H$\alpha$] (\AA) & $-10.44 \pm 0.22$ & $-16.79 \pm 0.18$ & $-110.04 \pm 0.29$ & $-23.91 \pm 0.25$ & 2\\
$v_{10}$[H$\alpha$] (km s$^{-1}$) & $210 \pm 14$ & $239 \pm 13$ & $367 \pm 16$ & $305 \pm 20$ & 2 \\
$\log(\dot{M}_{\mathrm{acc,H\alpha}} \mathrm{(M_{\odot}/yr)})$\tablenotemark{a} & $-10.9 \pm 0.4$ & $-10.6 \pm 0.4$ & $-9.3 \pm 0.4$ & $-9.9 \pm 0.4$ & 2 \\
EW[Li I 6707.8] (m\AA) & $<108$ & $<126$ & $<116$ & $<251$ & 2\\
Parallax (mas) & $12.1093 \pm 0.0629$ & $12.1604 \pm 0.0594$ & $12.6285 \pm 0.0742$ & $12.8064 \pm 0.0509$ & 1\\
$\mu_{\alpha} \cos \delta$ (mas/yr) & $33.351 \pm 0.084$ & $33.534 \pm 0.080$ & $-36.096 \pm 0.135$ & $-39.135 \pm 0.099$ & 1\\
$\mu_{\delta}$ (mas/yr) & $-5.459 \pm 0.118$ & $-3.629 \pm 0.112$ & $28.565 \pm 0.131$ & $23.582 \pm 0.117$ & 1\\
Radial velocity ($\mathrm{km s^{-1}}$) & $26.7 \pm 16.8$ & $29.8 \pm 16.8$ & $22.4 \pm 16.7$ & $20.5 \pm 16.8$ & 2\\
Excess ([W1]-[W3]) & \multicolumn{2}{c}{$1.146 \pm 0.034$} & \multicolumn{2}{c}{$1.468 \pm 0.027$} & 2\\
Excess ([W1]-[W4]) & \multicolumn{2}{c}{$3.093 \pm 0.078$} & \multicolumn{2}{c}{$2.888 \pm 0.043$} & 2\\
Stellar temperature & $\sim 2800$ & $\sim 2800$ & $\sim 3050$ & $\sim 3200$ & 4\\
%Hot disk temperature (K) & \multicolumn{2}{c}{} & \multicolumn{2}{c}{\nodata} & 2\\
Warm disk temperature (K) & \multicolumn{2}{c}{$300^{+11}_{-10}$} & \multicolumn{2}{c}{$279^{+7}_{-6}$} & 2\\
$L_{\mathrm{ir}}/L_{\star}$ & \multicolumn{2}{c}{$0.0262 \pm 0.0006$} & \multicolumn{2}{c}{$0.0178^{+0.0003}_{-0.0004}$} & 2\\
\enddata
\tablerefs{(1) \citet{2018AandA...616A...1G}. (2) This work. (3) \citet{2018ApJ...856...23G}. (4) \citet{2013ApJS..208....9P}.}
\end{deluxetable*}

\section{Analysis of New Peter Pan Disk Systems}
\label{sec:new_disks_analysis}

\subsection{Spectral types}

We used the PyHammer package \citep{2017ApJS..230...16K} to classify our four GMOS-S spectra. Given the proximity of the four targets to the Sun, we assumed solar metallicity for all four cases. We adopt spectral types of M6 for both components of J0446, and find that J0949 consists of an M4 and an M5. The four spectra, with comparison spectral standards, are shown in Figure \ref{fig:gmos_spectra}. 

\subsection{Age estimates and radial velocity measurements}

To confirm the membership of these specific objects in moving groups, we must measure the stellar heliocentric radial velocity (to ensure that its full kinematics match those of the group), and identify an independent age constraint.

We determined radial velocities for each of the four targets based on Gaussian fitting of four prominent lines (H$\alpha$, K I $\lambda$7669, Na I $\lambda$8183, and Ti I $\lambda$8435) to determine their observed central wavelengths. Uncertainties for each measurement reflect the 68\% confidence interval of the line profile as determined using \texttt{emcee} \citep{2013ascl.soft03002F}, the modest resolution of the spectrograph, and instrumental uncertainty due to telescope flexure across nights. These measurements are listed in Table \ref{table:new_disks}. Within the uncertainties, the measured radial velocities are consistent with membership in their assigned moving groups.

\begin{figure}
\begin{centering}
\includegraphics[width=0.5\textwidth]{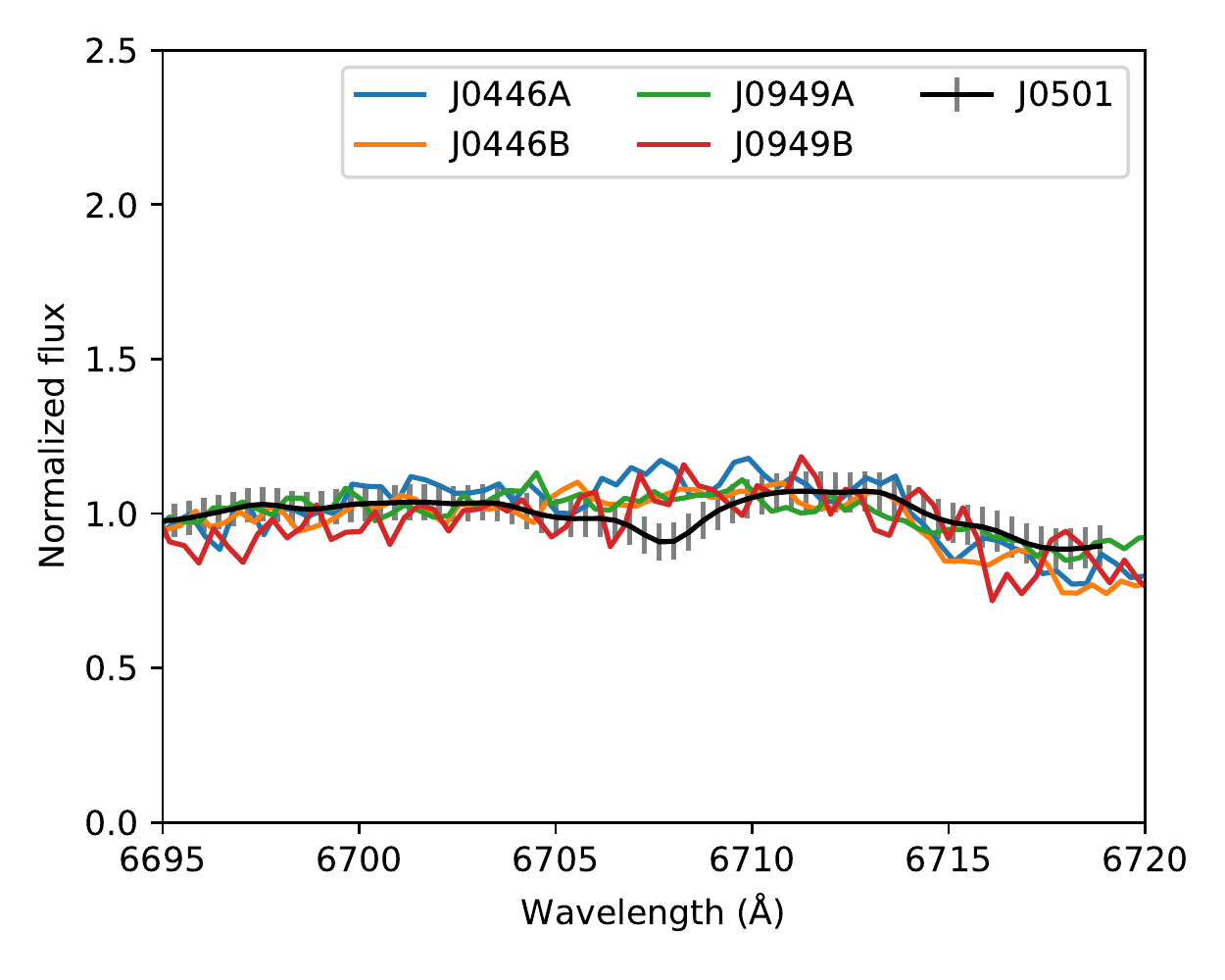}
\caption{GMOS spectra of the region around the Li I absorption line at 6707.8 \AA\ in four new Peter Pan disk candidates, compared with observations of J0501 \citep{2016ApJ...832...50B} smoothed to the resolution of the GMOS data. None of the four disk candidates shows significant absorption, though the rebinned spectrum of J0501, an M dwarf of similar age, suggests that lithium should be detectable.} 
\label{fig:lithium}
\end{centering}
\end{figure}

Figure \ref{fig:lithium} depicts the spectra for our four targets around the lithium feature at 6707.8 \AA, along with the spectrum of J0501 \citep[adopted from][]{2016ApJ...832...50B}, degraded to the resolution and rebinned to the dispersion of our data. As seen in Figure \ref{fig:lithium}, we do not detect clear evidence of lithium in our data, deriving upper limits of 0.11-0.25 \AA. Lithium absorption is often used as an age indicator \citep[and references therein]{2018MNRAS.476.3290M}; these upper limits are consistent with ages 40-45 Myr and our moving group assignments.

\begin{figure}
\begin{centering}
\includegraphics[width=0.5\textwidth]{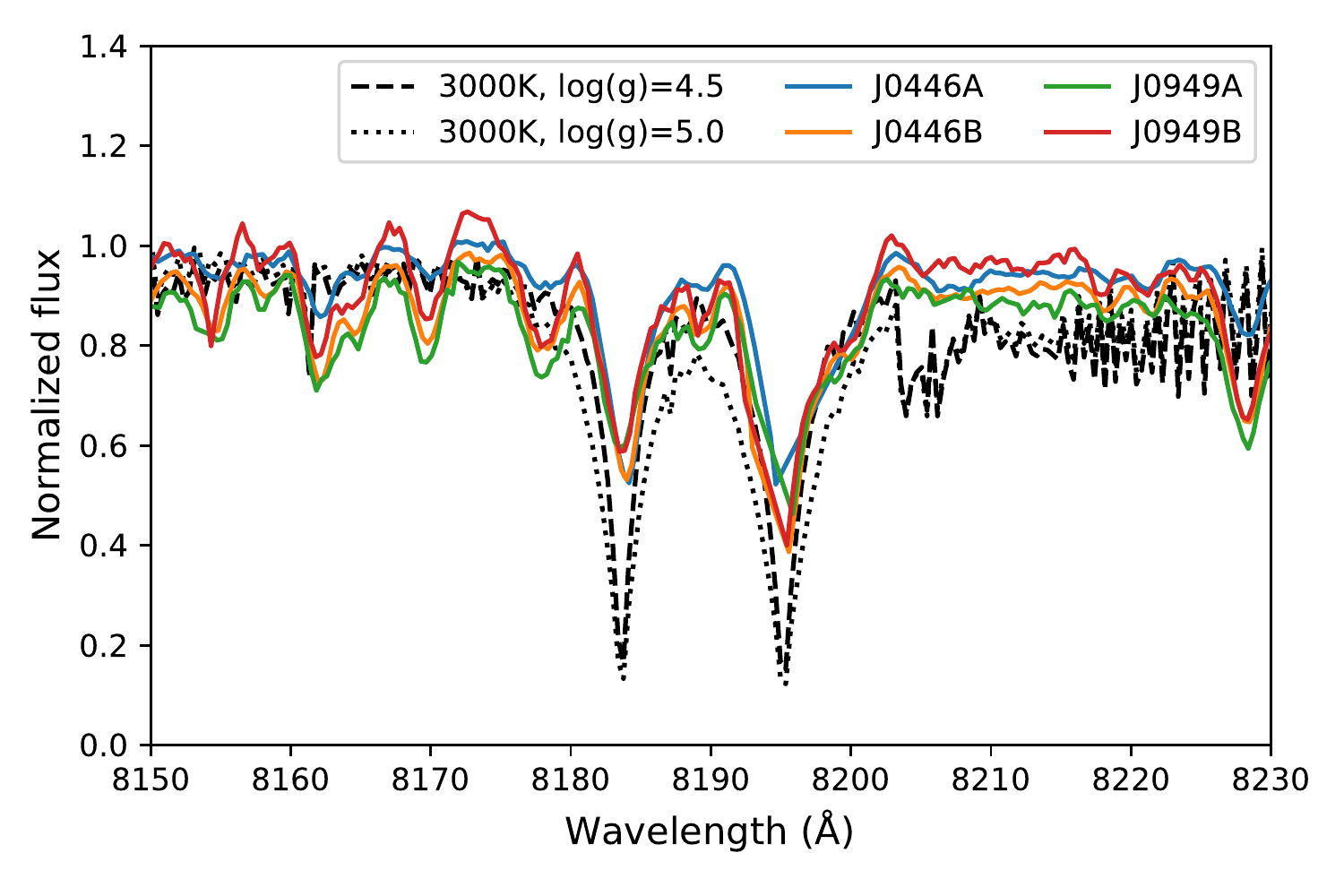}
\caption{GMOS spectra of the Na I 8200 \AA\ feature in four new Peter Pan disk candidates, compared with BT-Settl CIFIST models \citep{2015A&A...577A..42B} smoothed to the resolution of the GMOS data and shifted to the wavelength scale of the GMOS data. All four disk candidates exhibit Na I 8183 \AA\ absorption weaker than the $\log(g) = 5.0$ model, especially in the wing profiles. The line wing profiles for a $\log(g) = 4.5$ model are a better match, indicating weaker gravity than a typical dwarf and therefore youth.}
\label{fig:NaI_doublet}
\end{centering}
\end{figure}

Another indicator of youth is the depth of the Na I doublet at 8200 \AA\, which can be used to differentiate young cluster and moving group members from field stars and giants \citep{2012AJ....143..114S}. The 8195 \AA\ line of this feature falls on a bad pixel column in our data, making an equivalent width measure of the feature in our data unreliable. However, we can compare the observed line profiles of the 8183 \AA\ line to stellar models to qualitatively assess the line. As seen in Figure \ref{fig:NaI_doublet}, all four of our spectra are much shallower at 8183 \AA\ than a BT-Settl \citep{2015A&A...577A..42B} model of a main-sequence mid-M dwarf, suggesting that they have weaker gravity and are therefore younger than the typical field M dwarf.

\subsection{H$\alpha$ emission and accretion}

\begin{figure}
\begin{centering}
\includegraphics[width=0.5\textwidth]{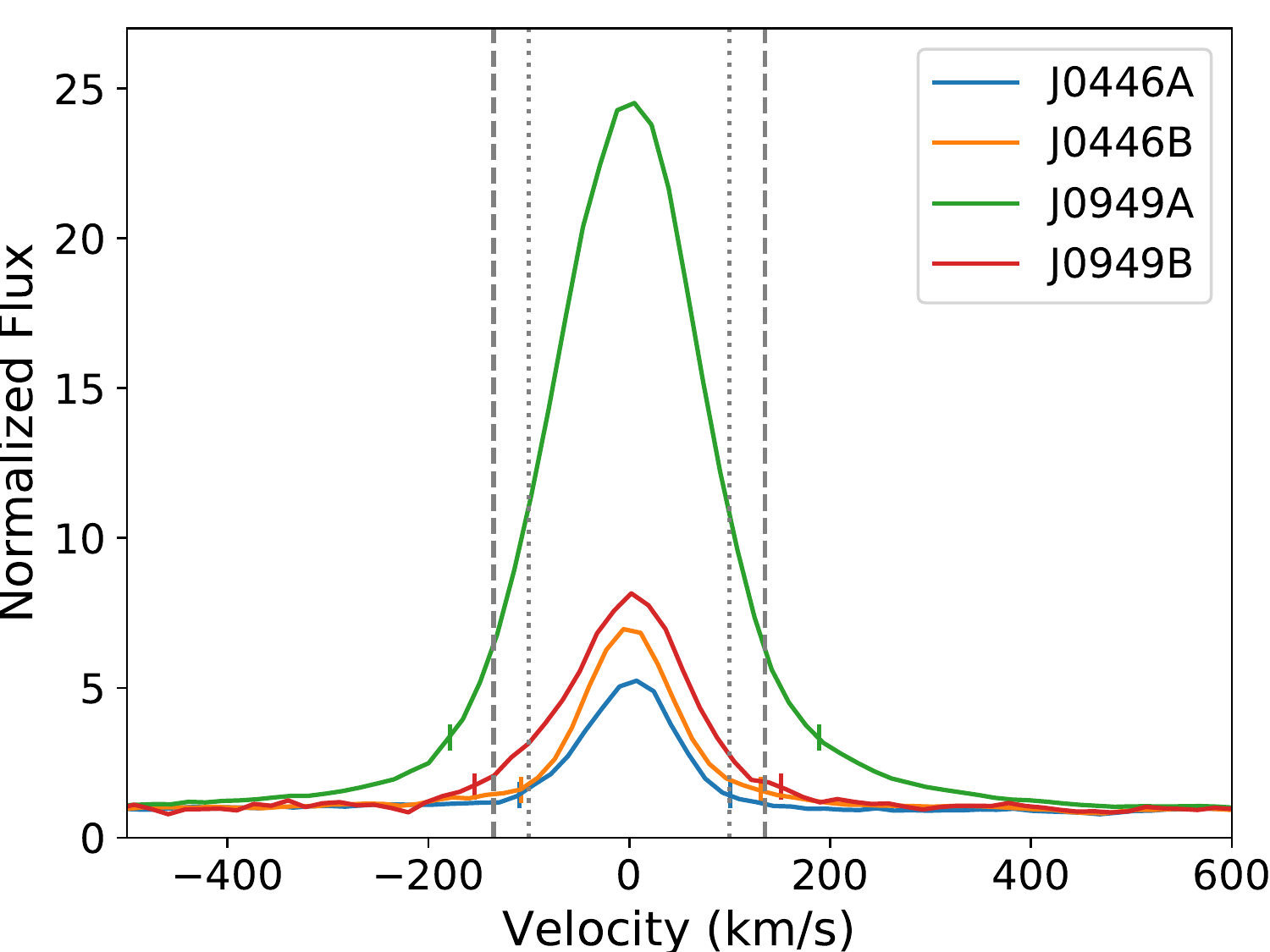}
\caption{GMOS velocity H$\alpha$ profiles of the four new Peter Pan disk candidates. Vertical markers indicate the velocity at which the profile is at 10\% of maximum. The dotted lines are the symmetric $v_{10}=200 \mathrm{km s^{-1}}$ criteria defined by \citet{2003ApJ...592..282J} and \citet{2004A&A...424..603N} The dashed line indicates the $v_{10} = 270 \mathrm{km s^{-1}}$ criterion defined by \citet{2003ApJ...582.1109W}.}
\label{fig:velocity_line_profiles}
\end{centering}
\end{figure}

Accretion of disk material from a circumstellar disk onto the host star is accompanied by excess continuum Balmer emission \citep[e.g.][]{2016ARA&A..54..135H} and enhanced line emission \citep[e.g.][]{1998AJ....116.2965M}. $U$- and $u$-band photometry of these systems has not yet been observed, so we focus on line emission diagnostics. We identify H$\alpha$ emission on all four of our candidates, as shown in Figure \ref{fig:velocity_line_profiles}. We computed the equivalent width of each line via direct integration of the observed spectrum after continuum fitting. We list these widths, ranging from -110 \AA $<$ EW[H$\alpha$] $<$ -10 \AA, in Table \ref{table:new_disks}. 

\citet{2009A&A...504..461F} proposed an equivalent width limit of $-18$ \AA\ for H$\alpha$ for a mid-M accretor, with many classical T Tauri stars exceeding this bound. While J0446A and J0446B do not meet the \citet{2009A&A...504..461F} limit, both J0949A and J0949B do. All four objects show an equivalent width larger in magnitude than the -7.56 angstroms exhibited by 2MASS J0501 \citep{2016ApJ...832...50B}, but only J0949A approaches the -125 \AA\ $<$ EW[H$\alpha$]$<$ -65\AA\ values observed for J0808.

Following \citet{2018MNRAS.476.3290M}, we estimated the velocity width at one-tenth maximum $(v_{10})$ by normalizing the line profile to a linear fit to the continuum at $\pm1000-500 \mathrm{km s^{-1}}$, and finding the numerical intercept of the line profile and the $f=v_{10}$ line (as shown in Figure \ref{fig:velocity_line_profiles}). \citet{2003ApJ...582.1109W} adopt a $v_{10} > 270 \mathrm{km s^{-1}}$ criterion for accretion across all M-types, while \citet{2003ApJ...592..282J} adopt a  $v_{10} > 200 \mathrm{km s^{-1}}$ for M5-M8 stars, and \citet{2004A&A...424..603N} adopts a $v_{10} > 200 \mathrm{km s^{-1}}$ criterion for M6-M8.5 stars. We adopt the \citet{2003ApJ...592..282J} criterion for the three stars with spectral types M5-M6, and adopt the \citet{2003ApJ...582.1109W} criterion for J0949A, which has a spectral type out of this range. Based on these criteria, all four objects appear to be accreting. Table \ref{table:new_disks} lists $v_{10}$ widths for the four stars. 

\begin{figure}
\begin{centering}
\includegraphics[width=0.5\textwidth]{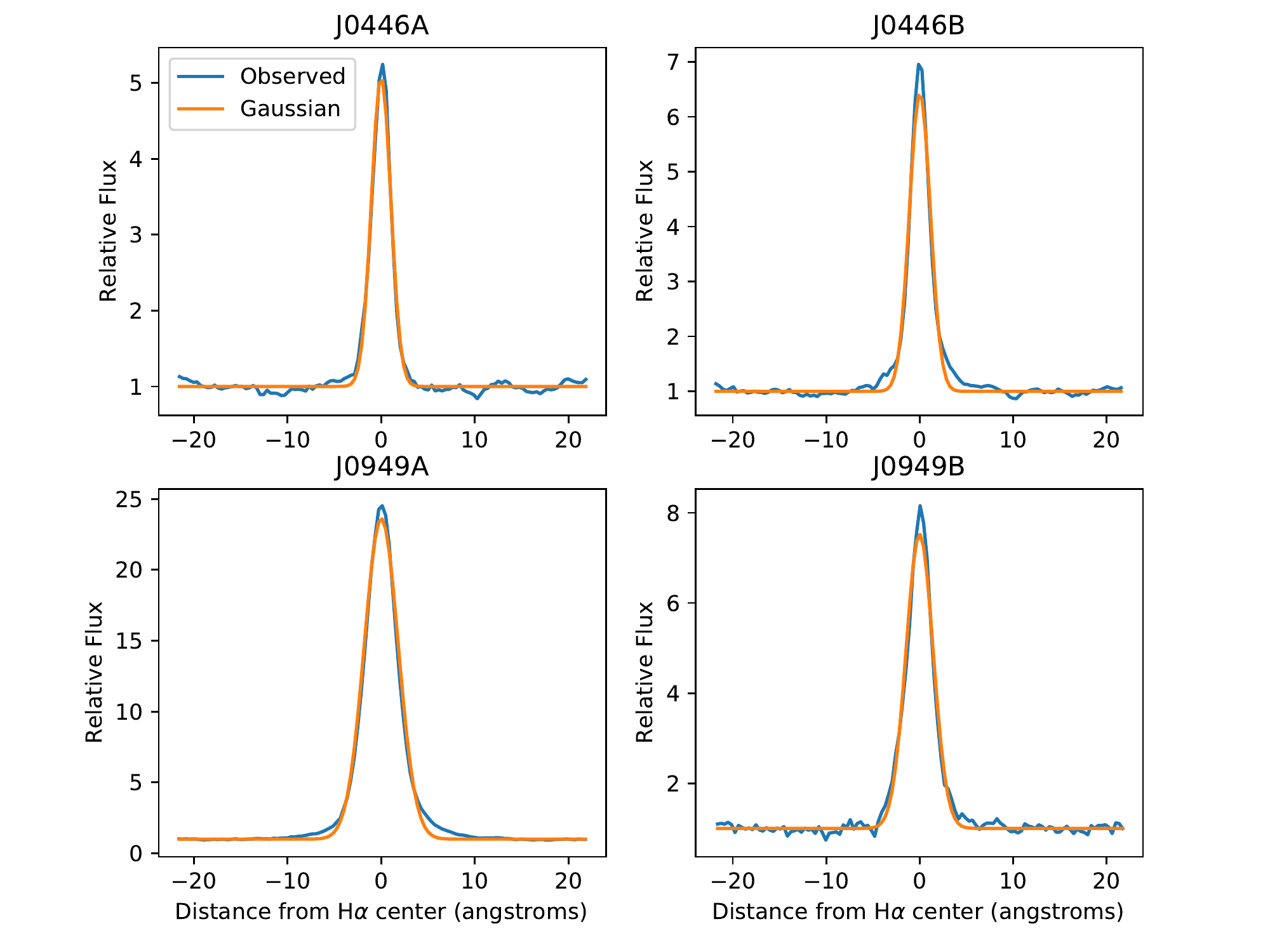}
\caption{Observed H$\alpha$ line profiles of the four targets, compared with best-fit Gaussians. Each example exhibits some qualitative asymmetry.}
\label{fig:lineprofiles}
\end{centering}
\end{figure}

Asymmetry of the line profile can also indicate whether emission is due to accretion rather than stellar activity, as infalling material should have some velocity shift relative to the host star \citep[e.g.][]{2016ARA&A..54..135H}. We present the H$\alpha$ profiles of the four targets in Figure \ref{fig:lineprofiles}, plotted against the best-fit Gaussian for each object to provide a symmetric reference to guide the eye. The line cores appear symmetric, once the asymmetric distribution of pixels across the line is accounted for. J0949B shows no obvious asymmetry. J0446A shows enhancement in the blue wing of the line compared to the red wing, though this may be due to the severe negative slope of the continuum in this range and imperfect continuum subtraction. Counter to this, J0949A exhibits an enhanced red wing of the profile compared to the blue wing. J0446B shows an asymmetric profile in the wings, similar to the asymmetric profile observed in 2MASS J05010082-4337102 (Boucher et al. 2016). None of the asymmetry observed here is to the degree of asymmetry exhibited by H$\alpha$ in typical young CTTSs \citep[e.g.][]{2016ARA&A..54..135H}; while there is apparent qualitative asymmetry, it is not quantitatively significant.  

We determined accretion rates using the \citet{2004A&A...424..603N} $v_{10,H\alpha}-\dot{M}_{acc}$ relation \citep[following][]{2018MNRAS.476.3290M,2016ApJ...832...50B}. Accretion rates (listed in Table \ref{table:new_disks}) range from $-10.8 < \mathrm{Log}(\dot{M}_{acc}) < -9.3$, suggesting lower limits on the mass accreted between $7 \times 10^{-4} \mathrm{M_{\odot}}$ and $0.02 \mathrm{M_{\odot}}$ in the stellar lifetime. While the lower bound of this range is consistent with ongoing accretion for $\sim 45$ Myr, the upper bound suggests that accretion may have recently increased.
{%\bf Take these accretion rates and multiply by the age of the stars: how much total mass did the star accrete during it lifetime under that assumption? Does that number sound reasonable as lower limit on the total accreted mass?  Or does it make you think that the accretion rate has recently increased?  -MJK}
However, these rates are likely higher than the true accretion rate, as the \citet{2004A&A...424..603N} relation was calibrated on younger (and therefore higher-radius) stars \citep[as noted by][]{2018MNRAS.476.3290M}.
%{\bf There is a well known relationship between accretiona rate an age for T Tauri stars, see Hartmann et al 1998, Muzerolle et al. 2000  (ADS links to classic papers in comments--but note there may be more recent version of this diagram!). How do the Peter Pan disks fit on these Mdot vs age diagrams?  What age is a typical disk whose accretion rate matches theirs?  -MJK}

%https://ui.adsabs.harvard.edu/abs/1998ApJ...495..385H/abstract
%https://ui.adsabs.harvard.edu/abs/2000ApJ...535L..47M/abstract

\section{Observations of Known Peter Pan Disks}
\label{sec:observations}

\subsection{High-Cadence Optical Imaging}

We obtained high-cadence optical photometry of J0808 over nine nights with the Tek2K CCD on the CTIO/SMARTS 0.9m telescope at Cerro Tololo Inter-American Observatory, from UT 2018 February 14-22, primarily to detect optical/UV flares \citep{1976ApJS...30...85L,2013ApJS..207...15K}. 
%{\bf What instrument did you use? -MJK}
Exposure times ranged 90-105 seconds exposures depending on sky transparency, to maximize signal-to-noise while maintaining as high a cadence as possible. We observed in the SDSS $g$ band to maximize contrast of the UV-bright flares with the underlying stellar photosphere while ensuring the faint source was detected in quiescence. To minimize time on-sky lost to readout, we observed in four-quadrant read-out, only exposing the inner $1024 \times 1024$ pixels. Data were reduced using standard IRAF procedures, and cosmic rays were removed using the \texttt{lacosmic} function \citep{2001PASP..113.1420V}. Relative differential photometric lightcurves were generated for each night using AstroImageJ \citep{2017AJ....153...77C}. We used an ensemble of six stars in the field of view as comparison stars for the target. To ensure that these six stars were non-variable, we analyzed light curves produced by differential photometry of each comparison star compared to the others in the ensemble; all observations in the light curves of these comparison stars are within 5$\sigma$ of the light curve median ($5\sigma = 1\%$ of the light curve median). Typical photometric uncertainties of the target are on the order of 2\%. An example image from the CTIO data is shown in Figure \ref{fig:Figure_new}. %{\bf How many percent is five sigma? -MJK}

\begin{figure*}
\plottwo{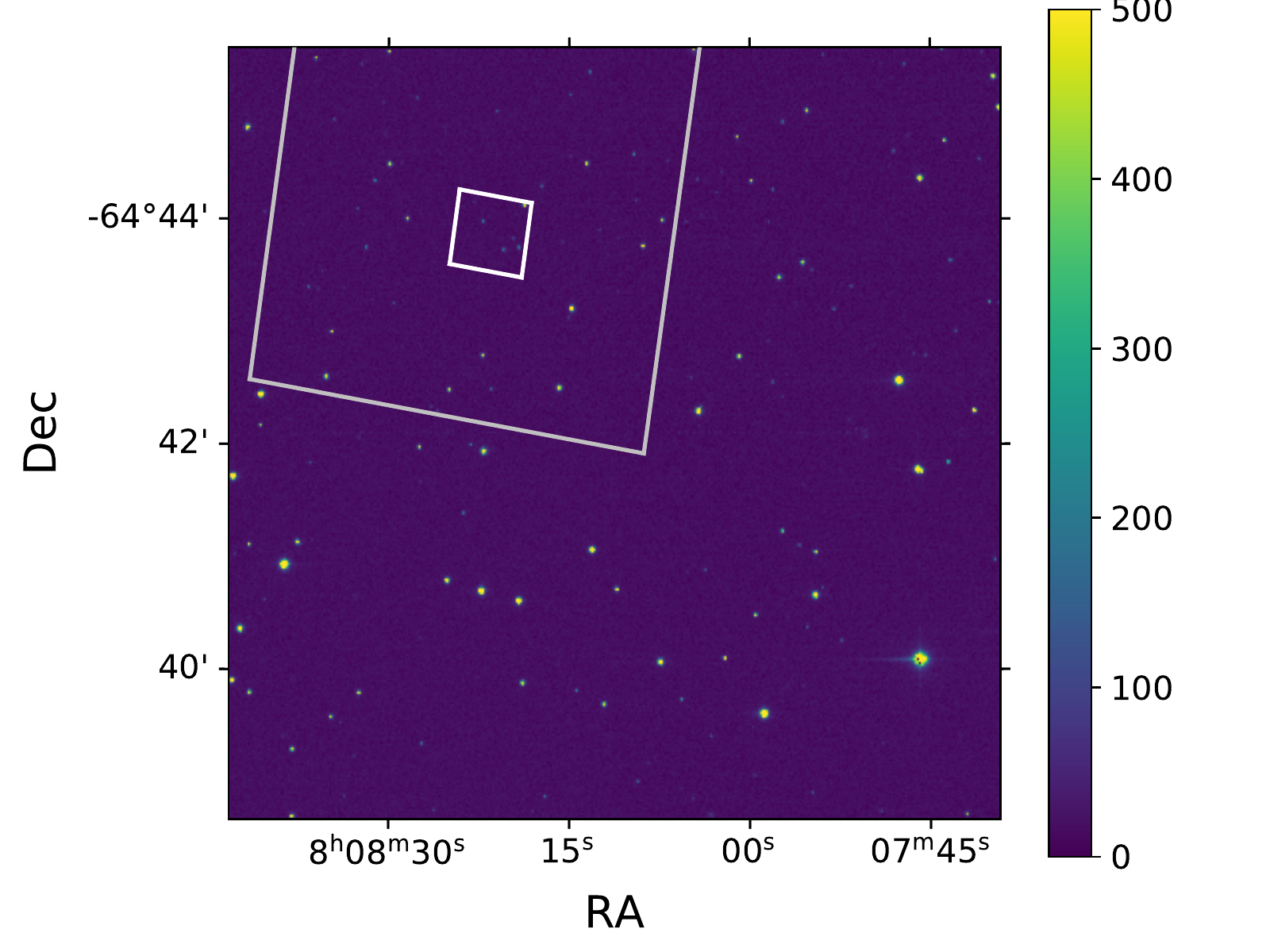}{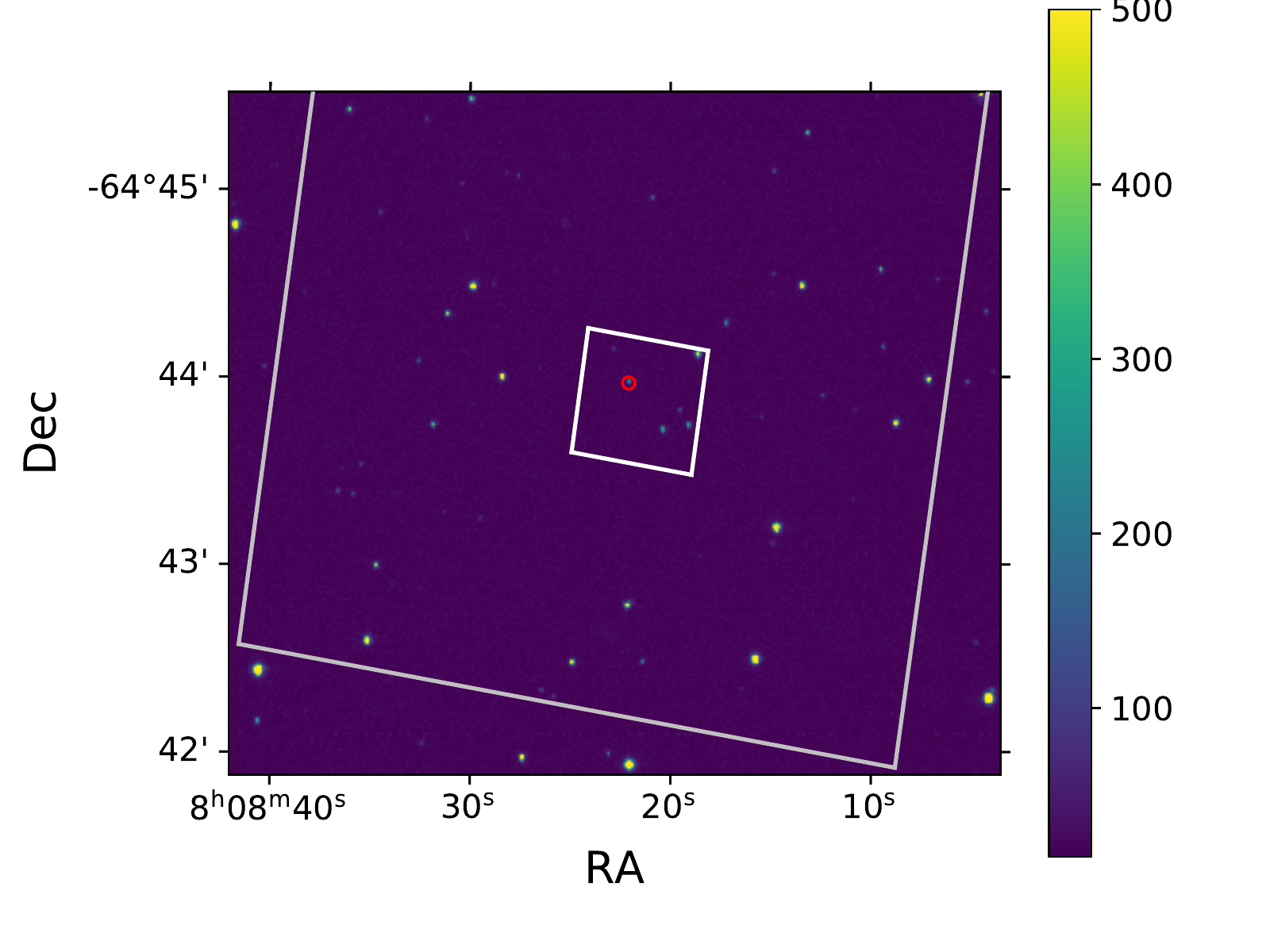}
\caption{\textit{Left}: An example image from the CTIO observing run. The TESS postage stamp (11 pixels $\times$ 11 pixels) is outlined in gray, while the aperture used to generate the lightcurve (4 pixels $\times$ 4 pixels) is outlined in white. \textit{Right}: A zoomed-in view of the TESS postage stamp. J0808 is highlighted by the red circle.} \label{fig:Figure_new}
\end{figure*}

Additionally, both J0808 and J0501, another Peter Pan disk candidate in the Columba system \citep{2016ApJ...832...50B}, were observed at two-minute cadence from 2018 October 18-November 15 by the Transiting Exoplanet Survey Satellite \citep[TESS;][]{2014SPIE.9143E..20R} during Cycle 1, Sector 4, as part of TESS GI Program G011148 (PI Kuchner). J0808 appeared in CCD 4 of Camera 4, while J0501 appeared in CCD 4 of Camera 3. The two-minute cadence data consist of $11\times11$ pixel subarrays, reduced with the Science Processing Operations Center (SPOC) pipeline, as described in \citet{2016SPIE.9913E..3EJ}. We use the two-minute Presearch Data Conditioning light curves from the SPOC pipeline, extracted from the apertures depicted in Figure \ref{fig:tess_aperture}. The aperture for J0808 includes contributions from four other nearby stars that are spatially resolved in the CTIO data, as shown in Figure \ref{fig:Figure_new}. We address the effect of these stars in Section \ref{sec:lightcurves}.

\begin{figure*}
\plottwo{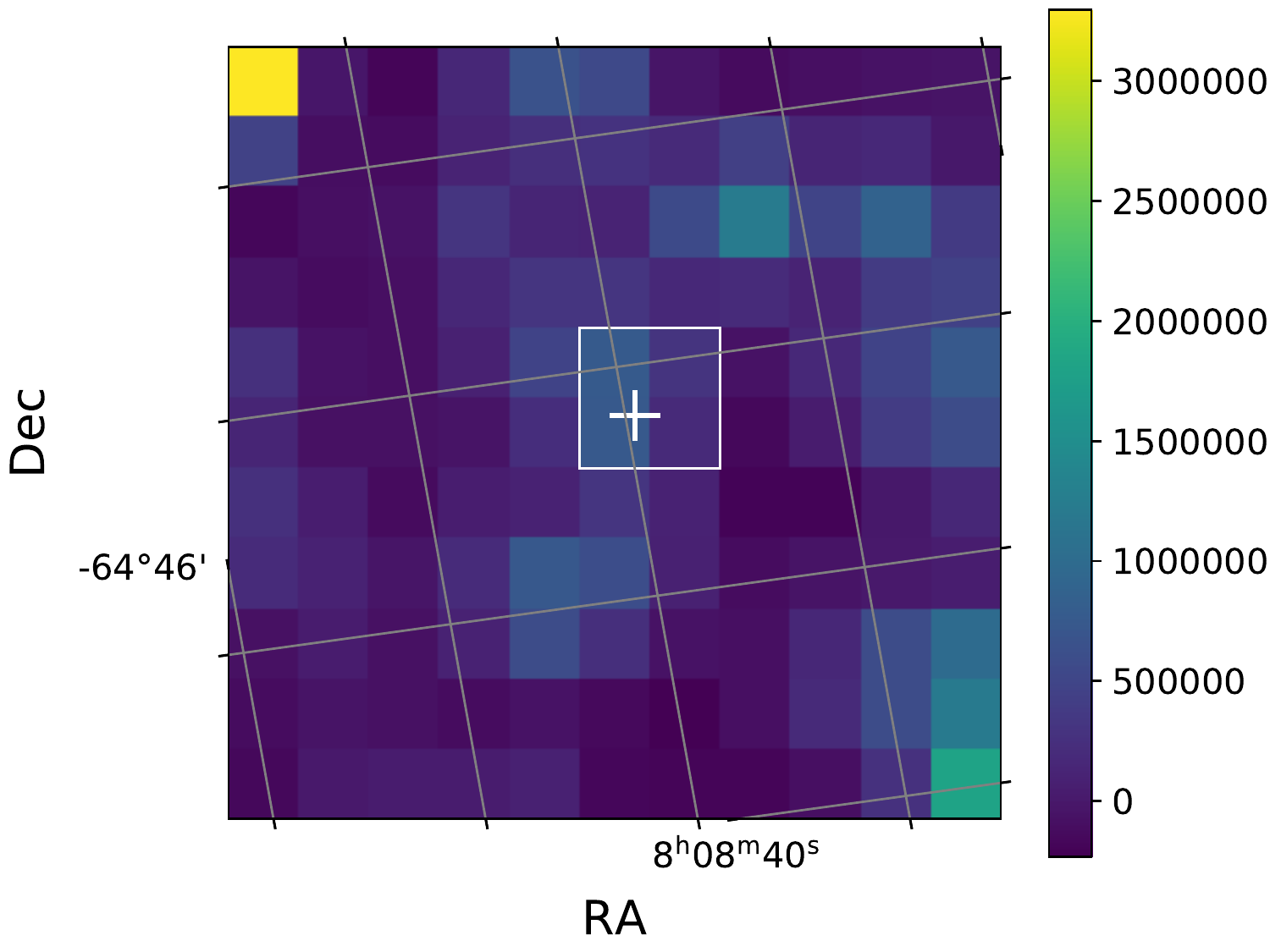}{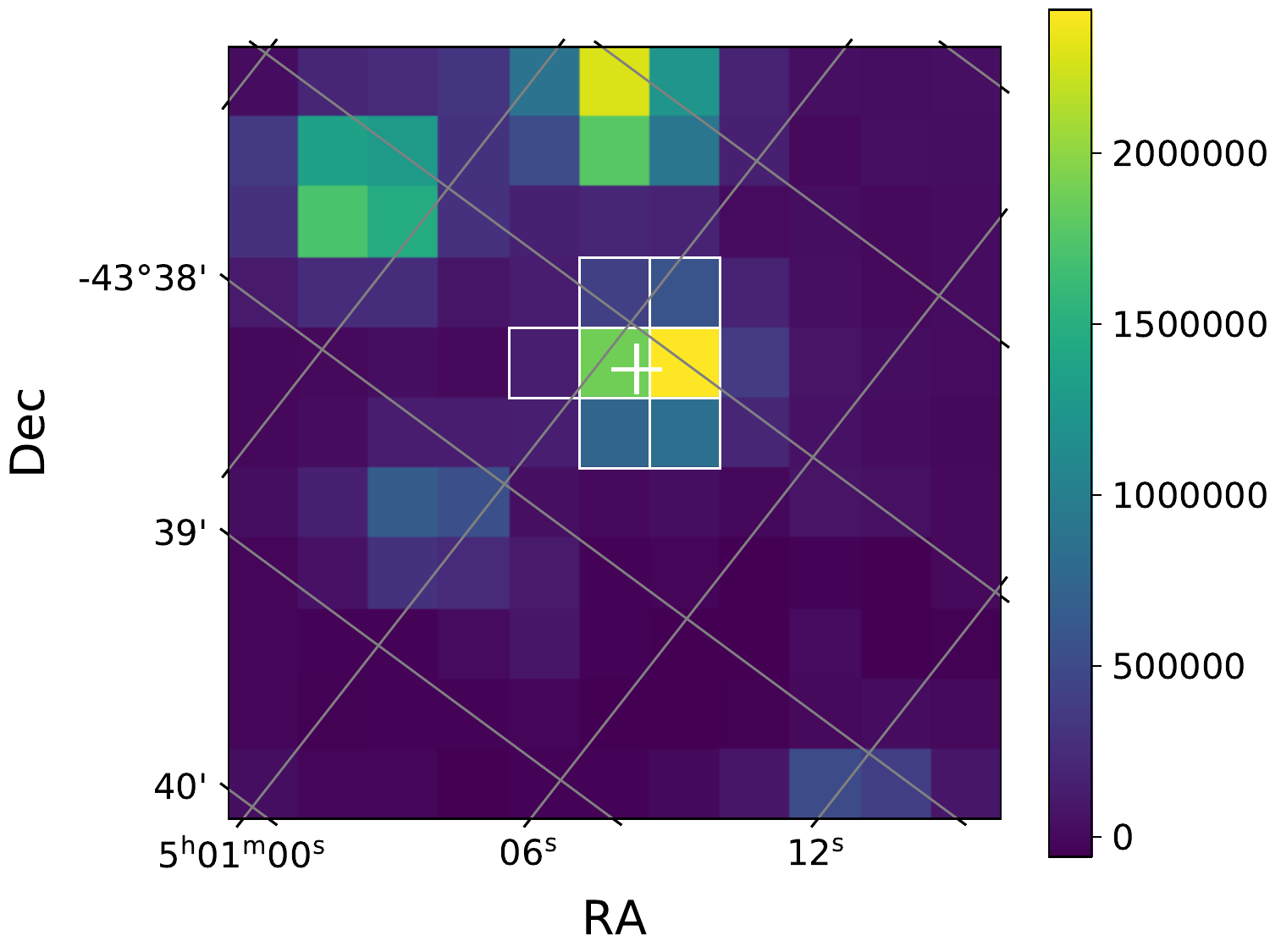}
\caption{The TESS target pixel files for J0808 (\textit{left}) and J0501 (\textit{right}) in Cycle 1, Sector 4. The PDC apertures are enclosed by white boxes. The \textit{Gaia} DR2 positions of the target stars are indicated by white crosses. Note the rotation in perspective for J0808 relative to the CTIO images.}
\label{fig:tess_aperture}
\end{figure*}

\subsection{Near-Infrared Spectroscopy}

We observed J0808 three times with the Astronomy Research using the Cornell Infra Red Imaging Spectrograph (ARCoIRIS) near-infrared spectrograph \citep{2014SPIE.9147E..2HS} on the 4m Victor Blanco Telescope at Cerro Tololo Inter-American Observatory. Observation dates, individual exposure times, and total integration times are listed in Table \ref{table:nearir_obs}. We observed multiple spectra of the target and a nearby A0V star (for use in telluric correction) using an ABBA pattern to correct for sky variability over the observation. We reduced the data using the ARCoIRIS extension of the \texttt{Spextool} package \citep{2004PASP..116..362C}, combined spectra using the \texttt{xcombspec} package \citep{2004PASP..116..362C}, and corrected for telluric lines using the \texttt{xtellcor} package \citep{2003PASP..115..389V}.

\begin{deluxetable}{lcc}
\tablecaption{Observing Data for Near-IR Spectroscopy with Blanco/ARCoIRIS \label{table:nearir_obs}}
\tablewidth{0pt}
\tablehead{ & \colhead{Number} & \colhead{Total} \\ 
\colhead{Date} & \colhead{Exposures} & \colhead{Integration Time (s)}}
\startdata
2017-11-04 & 12 & 540 \\
2018-03-01 & 12 & 540 \\
2018-03-02 & 12 & 720 \\
\enddata
\end{deluxetable}

\section{Variability in High-Cadence Optical Photometry}
\label{sec:lightcurves}

\subsection{Varying Morphologies in Different Bandpasses of J0808}

Figure \ref{fig:CTIO_lightcurve_J0808} depicts the CTIO light curve of J0808, while Figure \ref{fig:tess_lightcurve_J0808} depicts the TESS lightcurve of J0808.  Qualitatively, both light curves are variable at the $10-20\%$ level; however, neither shows any obvious periodicity. While the TESS light curve is more temporally complete than the CTIO light curve, there are no obvious flares akin to the one observed in the CTIO photometry. 

\begin{deluxetable}{lccl}
\tablecaption{Light Curve Variability Statistics \label{table:light_curve_stats}}
\tablewidth{0pt}
\tablehead{\colhead{Target/Observatory} & \colhead{$Q$} & \colhead{$M$} & \colhead{Classification}}
\startdata
J0808/CTIO & $0.637$ & $-0.602$ & Aperiodic burster \\
J0808/TESS & $0.85$ & $0.20$ & Aperiodic dipper \\
J0501/TESS & $0.157$ & $-0.005$ & Periodic symmetric \\
\enddata
\end{deluxetable}

\citet{2014AJ....147...82C} present a morphology classification system for young stellar objects based on space-based high-cadence light curves. We adapted this system to characterize our Peter Pan disk light curves, adapting each statistical metric $Q$ and $M$ to our data sample. These statistics for each of the three light curves we analyze are listed in Table \ref{table:light_curve_stats}. 

\begin{figure*}[htbp]
\includegraphics[width=\textwidth]{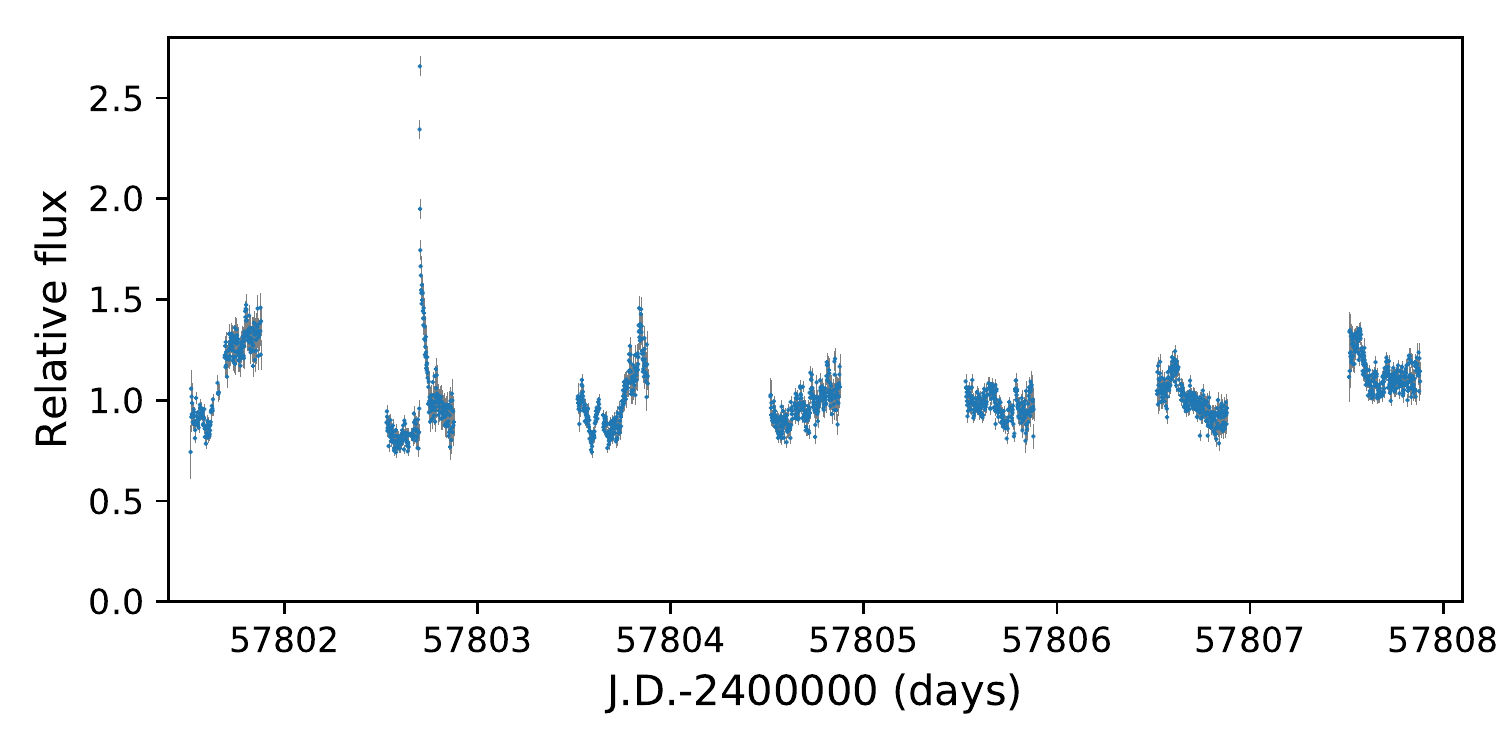}
\caption{The CTIO light curve of J0808 is characterized by short-duration aperiodic bursts, on timescales of half a night, and a large flare. Relative flux is calculated by dividing the sky-subtracted target star flux from the sky-subtracted flux of the comparison ensemble, and is normalized by dividing by the median.}
\label{fig:CTIO_lightcurve_J0808}
\end{figure*}

\begin{figure*}[htbp]
\includegraphics[width=\textwidth]{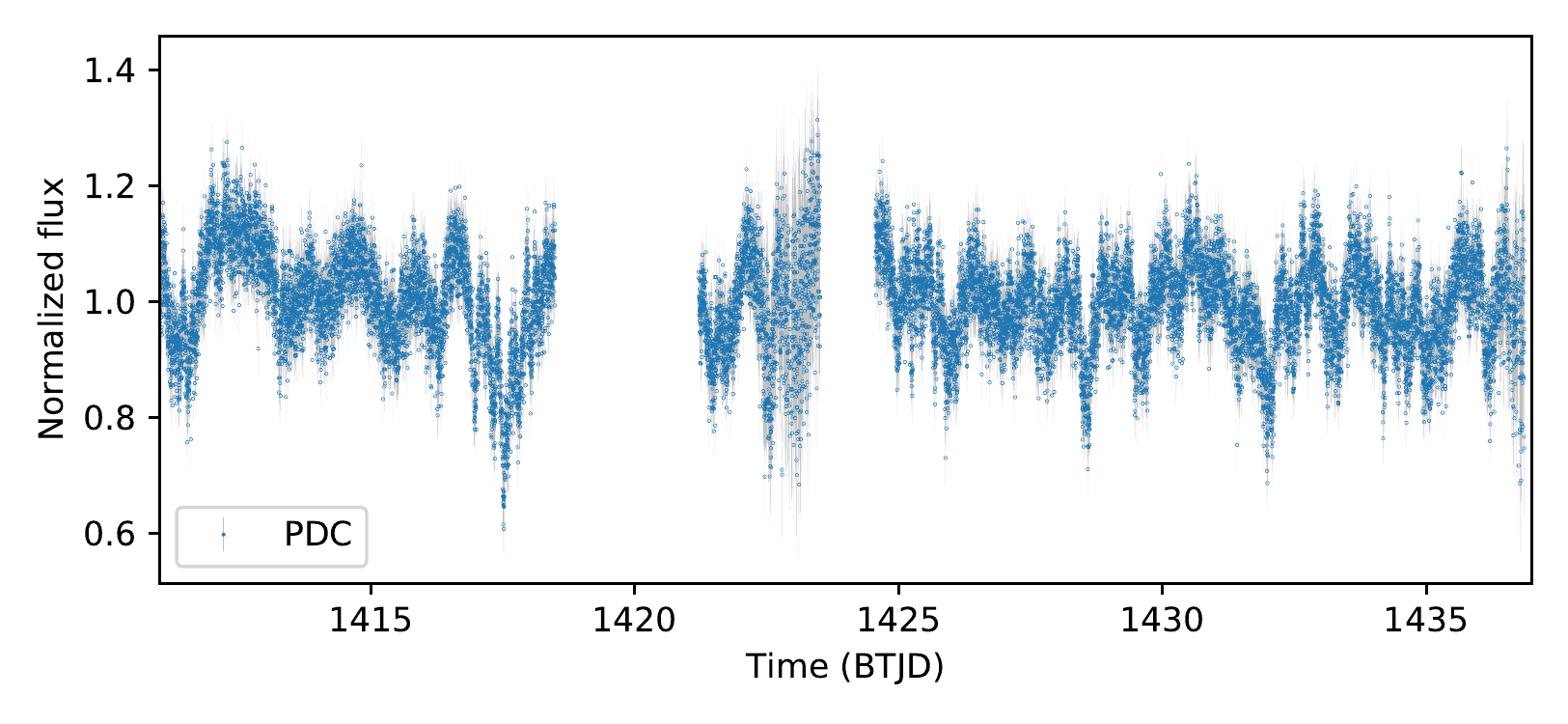}
\caption{The median-normalized TESS PDCSAP light curve of J0808 is characterized by aperiodic dipping on timescales of 0.5-2 days. Gaps are due to an instrument anomaly (TJD 1418.53-1421.21) and data download (1423.51-1424.55).}
\label{fig:tess_lightcurve_J0808}
\end{figure*}

Qualitatively, the CTIO light curve of J0808 exhibits stochastically-occurring, low-amplitude bursting events \citep{2014AJ....147...83S}, thought to correspond to accretion events. 
To quantify this, we applied a variation of the \citet{2014AJ....147...82C} $M$ statistic, which measures how asymmetric the light curve is with respect to reflection in the flux dimension (i.e. whether the light curve tends to burst, dip, or a mix of the two).
We smoothed the light curve for each night using a boxcar smoothing kernel of width 1.5 hours, and then identified $5\sigma$ outliers from the residual of the raw and smoothed light curves. We averaged the top and bottom 5\% of values from the outlier-removed light curve, subtracted this value from the median of the outlier-subtracted light curve, and divided by the root mean square uncertainty of the outlier-subtracted light curve. \citet{2014AJ....147...82C} define ``bursters'' as having $M < -0.25$, with bursts thought to be due to accretion events. ``Dippers'' are defined as having $M > 0.25$; the namesake dips are thought to be due to extinction by transiting dust in the disk. Objects with $M$ values between these limits are classified as ``symmetric.'' For the CTIO data, $M = -0.602$, consistent with our qualitative ``bursting'' assessment.  

Periodicity in the CTIO light curve is more difficult to assess, given the degeneracies that arise from the gaps in the ground-based light curve. Due to these gaps, we discarded the autocorrelation-function-based assessment of \citet{2014AJ....147...82C}, instead adapting it to use the Lomb-Scargle periodogram \citep{1976Ap&SS..39..447L,1982ApJ...263..835S}. We show the Lomb-Scargle periodogram for J0808 and neighboring stars within the TESS aperture in Figure \ref{fig:LS_J0808}. We see strong periods of 1.10 days and 0.53 days for J0808, likely aliases of each other introduced by the windowing function of nightly observations. Similar signals do not appear for the other stars in the aperture, and the periodogram of the combined-flux light curve of all four stars is dominated by the signal found in J0808.

\begin{figure*}
\plottwo{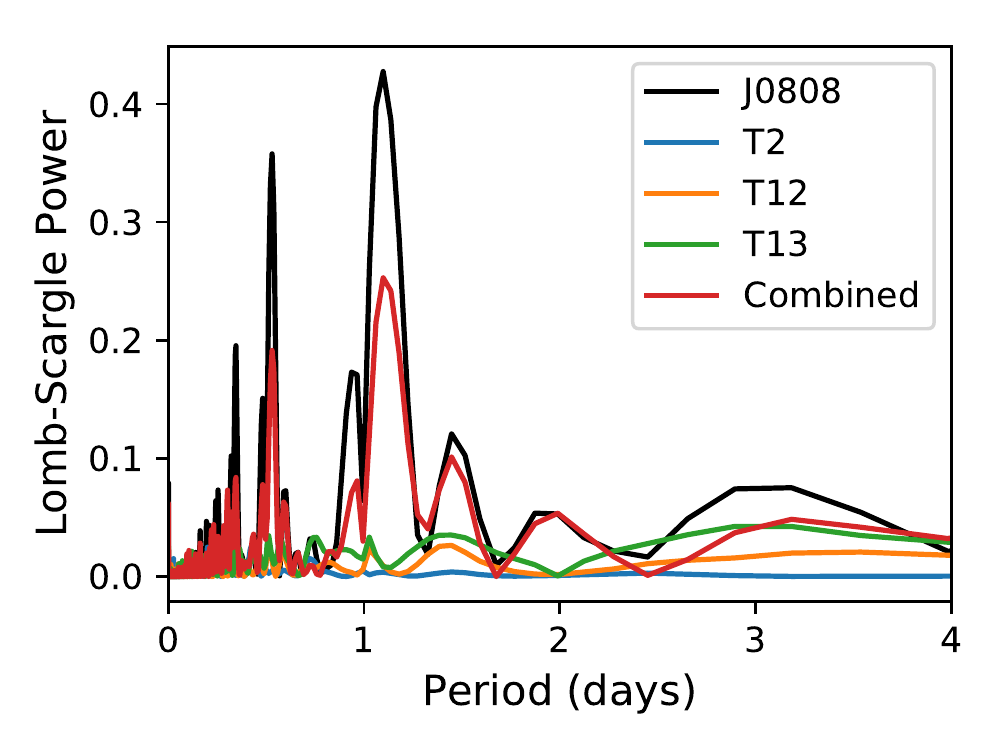}{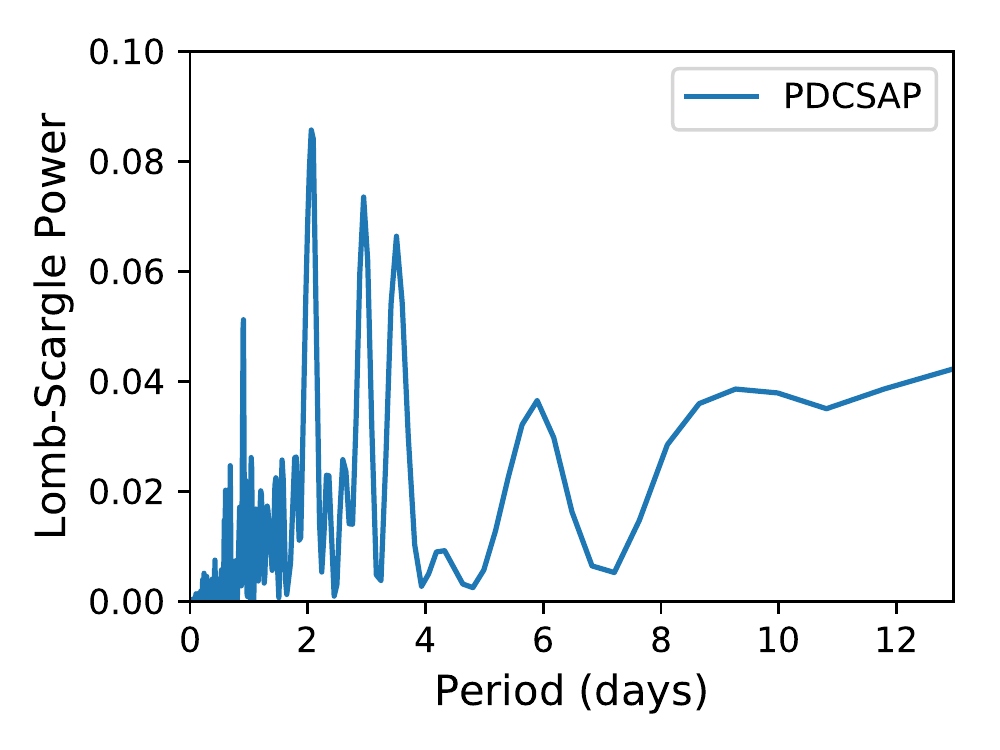}
\caption{\textit{Left}: Lomb-Scargle periodogram of CTIO lightcurves of J0808, three nearby stars within the TESS aperture, and the combined flux of the four stars. While J0808 shows two strong periods (likely aliases of each other from the observing window), none of the other three stars exhibit similar periodicity, suggesting that the variability is intrinsic to J0808 rather than a purely observational effect. The periodogram of the combined flux traces that of J0808, suggesting that the variability observed in the TESS data is due to J0808. \textit{Right}: The Lomb-Scargle periodogram for the TESS light curve of J0808 does not exhibit a signal at the periods found in the CTIO data, suggesting that the strength of this signal is due to the combination of the star's stochastic variability and the observing window.}
\label{fig:LS_J0808}
\end{figure*}

The phase-folded light curve assuming a 1.098-day period has a gap due to the nightly observing window, so we adopt the 0.531-day period to derive the $Q$ statistic from \citet{2014AJ....147...82C}, which measures how periodic a light curve is by determining the fraction of variability determined by the dominant periodic signal. With this period, we find $Q = 0.637$ for a 0.531-day period, an aperiodic variation per \citet{2014AJ....147...82C}. The most likely scenario is that the Lomb-Scargle periodogram is picking up this aperiodic variability. 

The TESS light curve does not exhibit these morphological characteristics. We qualitatively assess the light curve, shown in Figure \ref{fig:tess_lightcurve_J0808}, as an aperiodic dipper, with dips on the timescale of half a day to several days. The Lomb-Scargle periodogram for the target (Figure \ref{fig:LS_J0808}) has no peaks with power greater than 0.1, further bolstering the case for aperiodicity. To quantify the level of (a)periodicity, we calculate the $Q$ statistic using the strongest period in the light curve's Lomb-Scargle periodogram, which yields $Q=0.85$. We find $M=0.20$; \citet{2014AJ....147...82C} classify this as symmetric, but closer to the dipper end ($M>0.25$) than the burster end. 

We observed one large classical flare during the CTIO observations, in the second night depicted in Figure \ref{fig:CTIO_lightcurve_J0808}, using the PyVan software \citep{2019arXiv190303240L} for identifying and characterizing flares in unevenly-sampled photometry. The flare has a peak amplitude of $2.378\times$ the median of the full light curve, and a duration of 0.08 days. Using the empirical flare template for GJ 1243 from \textit{Kepler} \citep{2014ApJ...797..122D}, we estimate a flare equivalent duration of $\sim955$ s. Traditionally flare energies are calculated from the equivalent duration \citep[e.g.][]{1972Ap&SS..19...75G}; however, this requires an accurate determination of the stellar luminosity during the flare. Given that the star exhibits variability at the 20\% level likely due to ongoing accretion, it is impossible to confidently assess the underlying luminosity of the star during the flare without additional data (e.g. simultaneous spectroscopy), making an accurate energy determination impossible with the data in-hand. 

The TESS data do not reveal any obvious flares, likely because the TESS bandpass is redward of the peak flare brightness, and the SNR of the TESS data (typically $\sim 20-25$) is low compared to the CTIO data. To quantify this, we derive a conversion to estimate the flux increase the flare observed with CTIO would produce in our TESS data, using known standard flare characteristics and the CTIO observations. We adopt a 3100K, log($g$)=4.5 BT-Settl \citep{2015A&A...577A..42B} model to represent J0808 \citep[per][]{2018MNRAS.476.3290M}, and a 10,000K blackbody to represent the flare \citep{2013ApJS..207...15K}. We then determine the filling factor $a$ such that 

\begin{equation}
   aF_{g,\mathrm{flare}}+F_{g,\star} = 3.378F_{g,\star},
\end{equation}

\noindent
where $F_{g,\star}$ is the stellar flux in the Sloan $g$ band, $F_{g,\mathrm{flare}}$ is the estimated Sloan $g$-band flux of the adopted 10,000K blackbody, and $3.378F_{g,\star}$ is the observed flux at the flare peak relative to the median of the light curve.

\begin{figure*}[htbp]
\includegraphics[width=\textwidth]{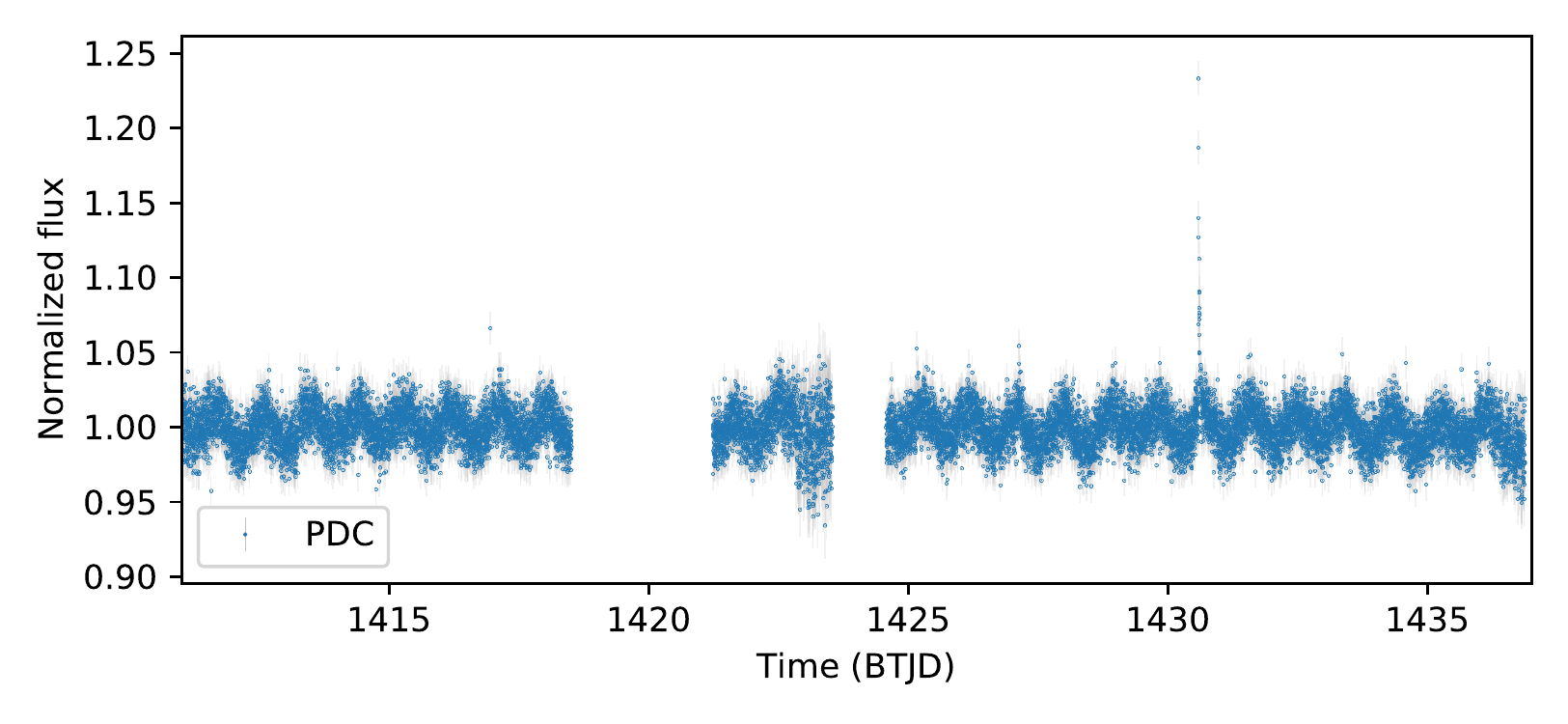}
\caption{The TESS PDCSAP light curve of J0501 is dominated by a periodic oscillation characteristic of a persistent starspot or complex of starspots. We also note a flare peaking at 1430.561 days.}
\label{fig:tess_lightcurve_J0501}
\end{figure*}

\begin{figure}
\begin{centering}
\includegraphics[width=0.5\textwidth]{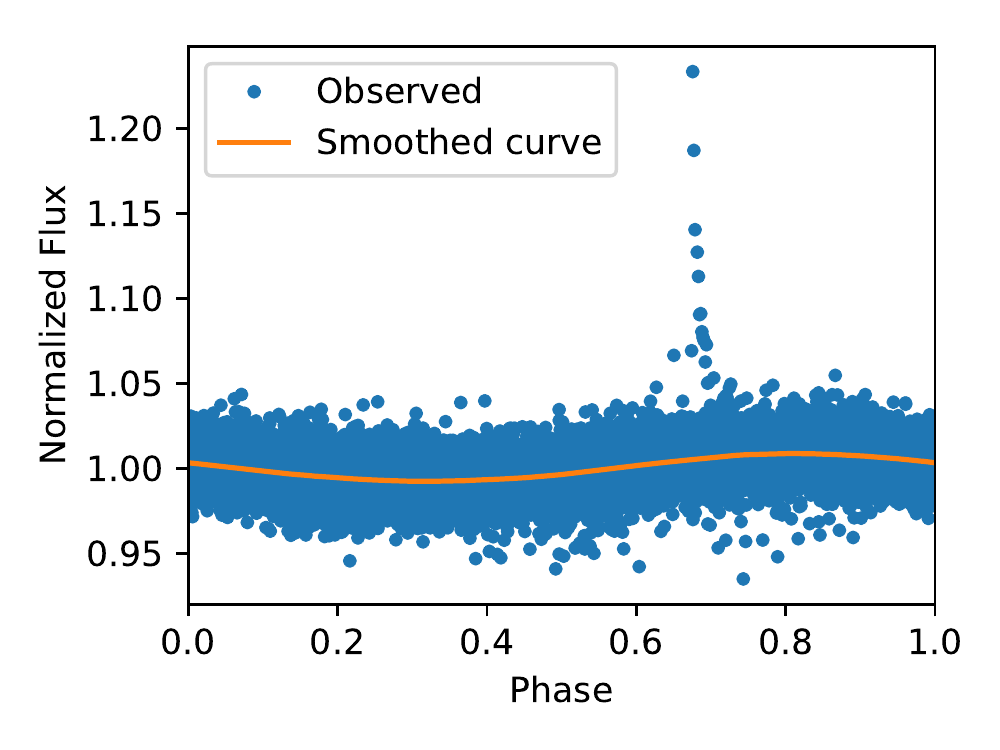}
\caption{The phase-folded light curve for J0501 shows clear sinusoidal modulation with a period of 0.907 days.}
\label{fig:phase_folded_lightcurve_J0501}
\end{centering}
\end{figure}

We then inverted this equation in the TESS bandpass, using the value of $a$ from the $g$ band to solve for the relative flux increase with TESS. We find that the CTIO flare would produce a 16.5\% increase in flux observed with TESS at the flare's peak, a 3.33$\sigma$ increase above the light curve median flux based on the light curve's RMS uncertainty. This is the most likely explanation for the lack of identifiable flares in the TESS data despite its increased temporal coverage; the TESS bandpass is red enough relative to the peak brightness of the flare that even large flares do not rise to the level of significance, especially on a faint target with correspondingly decreased signal-to-noise.

We also estimated the peak brightness in Sloan $g$ of a flare with peak brightness 5$\sigma$ above the median TESS flux. Such a flare would have a peak brightness $\sim 4.6\times$ the observed median with CTIO.

\subsection{Periodicity and a Flare on J0501}
\label{sec:tess_J0501}

Figure \ref{fig:tess_lightcurve_J0501} shows the TESS light curve for J0501. Unlike the light curve for J0808, J0501 exhibits both clear periodicity and an obvious flare. The $Q$ and $M$ statistics bear this out: $Q=0.157$, on the periodic/quasi-periodic boundary, while $M=-0.005$, nearly perfectly symmetric. The period of J0501 is 0.906 days, and the phase-folded light curve (shown in Figure \ref{fig:phase_folded_lightcurve_J0501}) exhibits a shape consistent with a persistent starspot or complex of starspots, as seen on the active M dwarf GJ 1243 with \textit{Kepler} \citep{2014ApJ...797..121H,2015ApJ...806..212D}

We estimate a stellar radius for J0501 of $0.32 R_{\odot}$, following the $J$-magnitude-based relation of \citet{2013ApJS..208....9P}.\footnote[2]{We adapt the technique from \citet{2013ApJS..208....9P} to use the 2015 IAU definition of $M_\mathrm{bol,\odot} = 4.74$, as was done in \citet{2018MNRAS.476.3290M}.} Using this, the observed TESS period, and the $v \sin i = 11 \mathrm{km s^{-1}}$ found by \citet{2016ApJ...832...50B}, we estimate an inclination angle for the system of $\sim38^{\circ}$. While not face-on, this indicates an inclination angle favorable for potential spatial resolution of the disk.

In addition to the rotation period, we identify one flare in the TESS data using PyVan and visual inspection. Fitting the \textit{Kepler} empirical flare template \citep{2014ApJ...797..122D} to it, we find a rise time of 385 seconds, and a duration of 106 minutes, yielding an equivalent duration of $11.6 \pm 1.6$ minutes.

\section{Accretion and Excess Detection in Near-Infrared Spectroscopy}
\label{sec:spectra}

\begin{figure}
\begin{centering}
\includegraphics[width=0.5\textwidth]{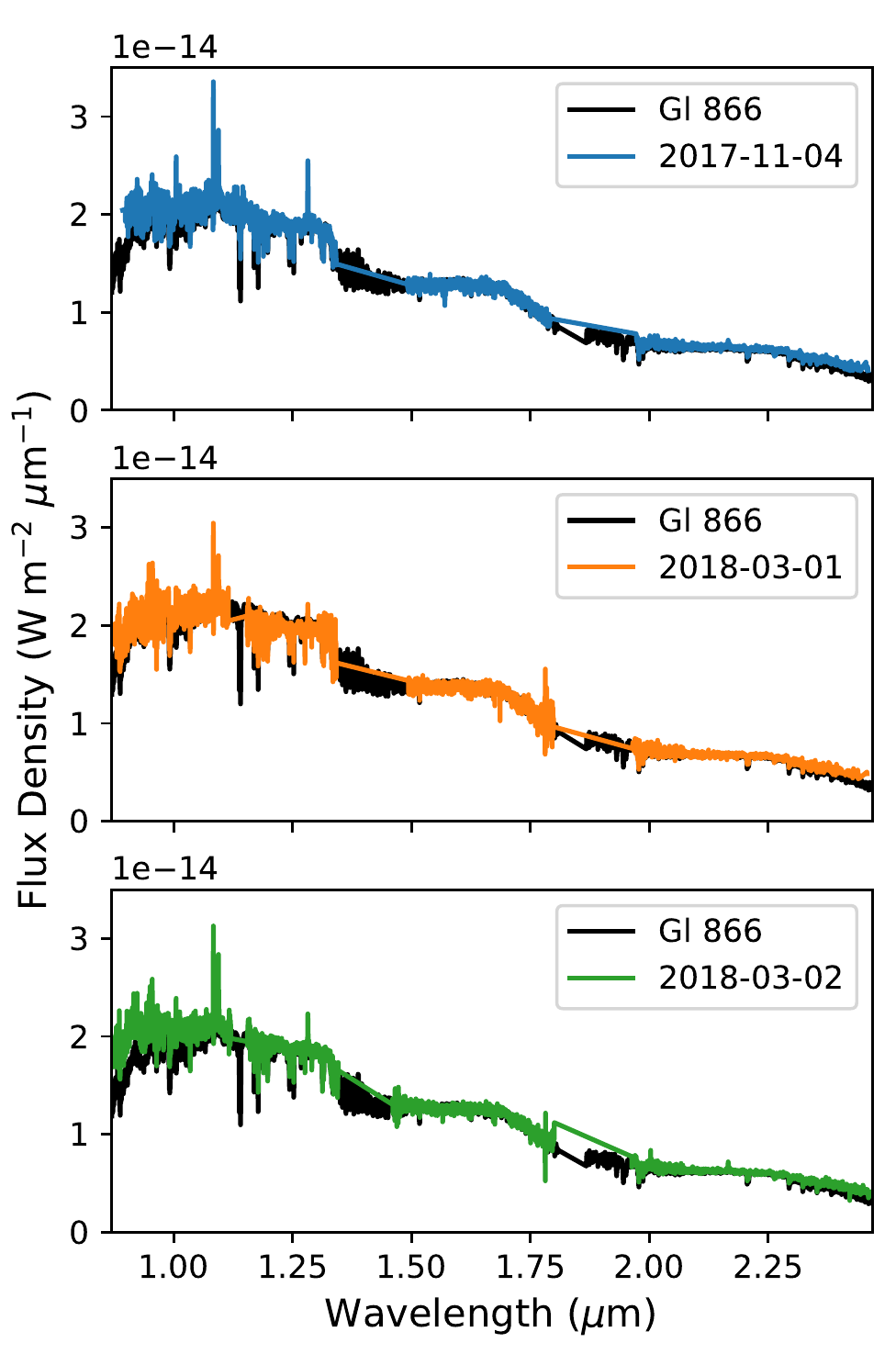}
\caption{Near-infrared spectra of J0808 from Blanco/ARCoIRIS, compared with Gl 866, a library M5V star used as a template. \label{fig:fullspectra}}
\end{centering}
\end{figure}

Figure \ref{fig:fullspectra} shows all three observations of J0808, compared with the SpeX library spectrum of Gl 866ABC \citep{2009ApJS..185..289R}, a typical M5V star that we use as a comparison template based on the M5V spectral type from \citet{2018MNRAS.476.3290M}. This comparison spectrum overall shows a good match to the J0808 spectra. However, there are significant deviations of interest in two places: J0808 exhibits an excess in its $K$-band spectra, and exhibits time-variable hydrogen emission.

\subsection{K-band Excess}

\begin{figure}
\begin{centering}
\includegraphics[width=0.5\textwidth]{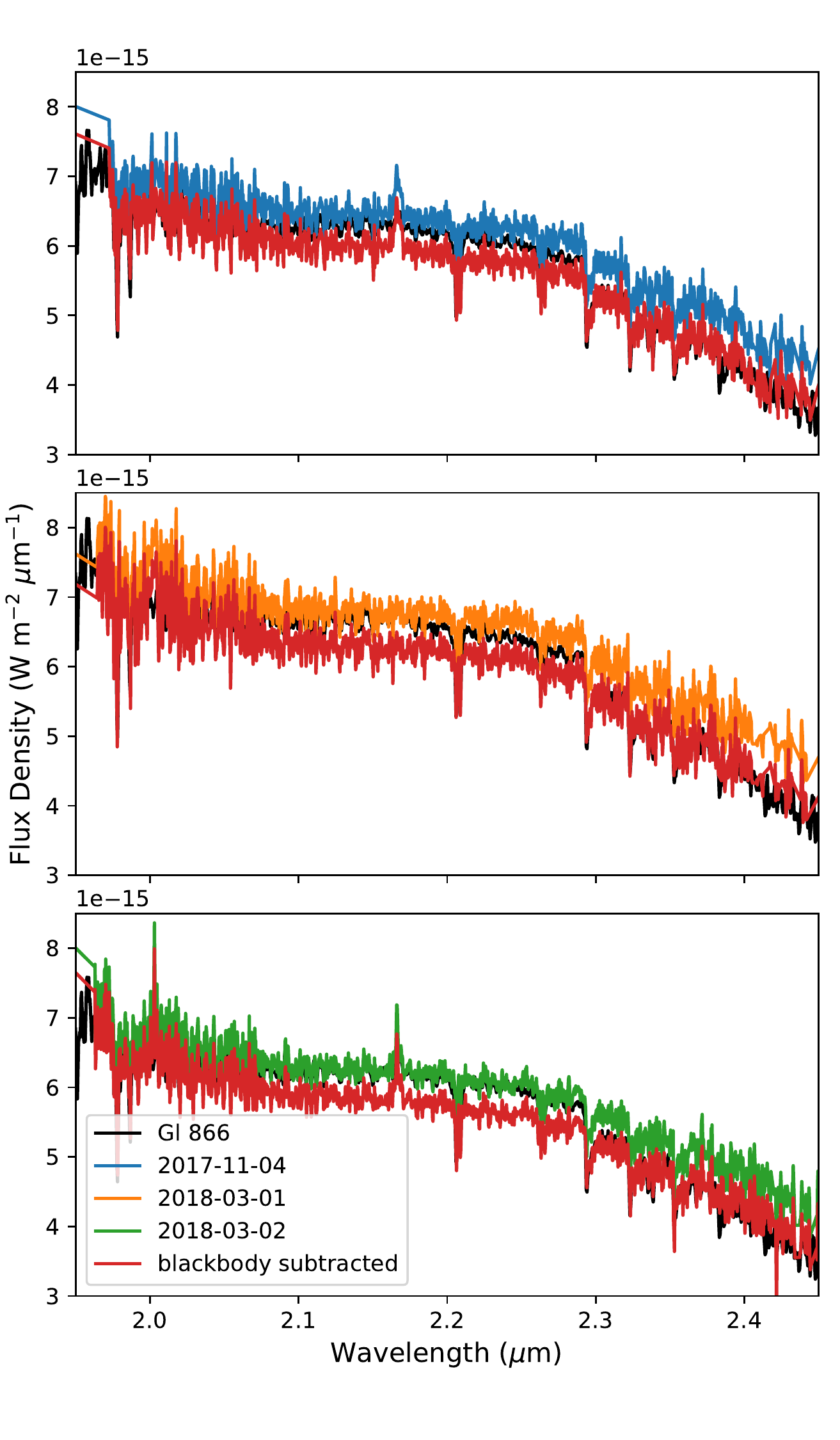}
\caption{K-band spectra of J0808 from Blanco/ARCoIRIS, compared with Gl 866 (black). The red line in each panel represents the observed spectrum with a 1071K blackbody with $L_{\mathrm{ir}}/L_{\star} = 0.11$ subtracted off. While the blackbody-subtracted spectrum matches Gl 866 well at wavelengths longer than 2.30 $\mu$m, the observed spectrum matches the template more closely at shorter wavelengths. \label{fig:Kspectrum}}
\end{centering}
\end{figure}

All three $K$-band spectra of the system (Figure \ref{fig:Kspectrum}) show a continuum level in excess of the expected continuum of an M5V star. 
\citet{2018MNRAS.476.3290M} identified correlated variability in the W1 and W2 single-exposure photometry for J0808, with a best-fit blackbody of $1071 \pm 103$ K and fractional infrared luminosity $L_{\mathrm{ir}}/L_{\star}$ ranging from 0 to 0.11, based on the observed variability. Our observations are consistent with detecting the short-wavelength end of this excess. The 1071 K blackbody at 0.11 $L_{\star}$ is a good fit to the excess beyond 2.30 $\mu$m in all three epochs; however, contributions from this blackbody do not appear at shorter wavelengths, suggesting a potential disk truncation radius. We do not detect any variability between observations above the continuum noise; however, the time series W1/W2 photometry has exhibited similar fluxes on 24-hour timescales. 

\subsection{Hydrogen Emission Variability}

%\begin{rotate}
\begin{deluxetable*}{lcccccccc}
\tablecaption{Hydrogen Emission Characteristics in Near-IR Observations of J0808 \label{table:hydrogen}}
\tablewidth{0pt}
\tablehead{\colhead{Date} & \multicolumn{3}{c}{Pa $\beta$} & \multicolumn{3}{c}{Br $\gamma$} & \colhead{Accretion} & \colhead{Accretion} \\ & \colhead{EW} & \colhead{Line Flux} & \colhead{Accr. Lum.} & \colhead{EW} & \colhead{Line Flux} & \colhead{Accr. Lum.} & \colhead{Luminosity} & \colhead{Rate} \\ & \colhead{\AA} & \colhead{$\mathrm{10^{-18} W m^{-2}}$} & \colhead{$\log L_{\mathrm{acc}}/L_{\odot}$} & \colhead{\AA} & \colhead{$\mathrm{10^{-18} W m^{-2}}$} & \colhead{$\log L_{\mathrm{acc}}/L_{\odot}$} & \colhead{$\log L_{\mathrm{acc}}/L_{\odot}$} & \colhead{$\log(\dot{M}_{\odot} yr^{-1})$} }
\startdata
2017-11-04 & $3.96 \pm 0.15$ & $7.48 \pm 0.30$ & $-3.6 \pm 1.1$ & $4.61 \pm 0.25$ & $2.95 \pm 0.17$ & $-3.2 \pm 1.4$ & $-3.4 \pm 1.1$ & $-10.8 \pm 1.3$ \\
2018-03-01 & $0.69 \pm 0.23$ & $1.36 \pm 0.45$ & $-4.7 \pm 1.3$ & $<0.98$ & $<0.66$ & $<-6.7$ & $-4.7 \pm 1.3$ & $-12.0 \pm 1.7$\\
%2018-03-01 & $0.69 \pm 0.23$ & $1.36 \pm 0.45$ & $-4.7 \pm 1.3$ & ... & ... & ... & $-4.7 \pm 1.3$ & $-12.0 \pm 1.7$\\
2018-03-02 & $1.08 \pm 0.21$ & $2.06 \pm 0.41$ & $-4.4 \pm 1.3$ & $4.02 \pm 0.27$ & $2.50 \pm 0.18$ & $-3.2 \pm 1.4$ & $-3.9 \pm 1.2$ & $-11.2 \pm 1.2$\\
\enddata
\end{deluxetable*}
%\end{rotate}

We determined equivalent widths and line luminosities for the Pa $\beta$ and Br $\gamma$ emission lines in all three spectra. From these we estimated the accretion luminosity, using the $\log(\mathrm{Br} \gamma)$ relation from \citet{1998AJ....116.2965M}. We use the $\log(\mathrm{Pa} \beta)$ relation from \citet{2004A&A...424..603N} rather than the similar relation from \citet{1998AJ....116.2965M} because the former incorporates data from objects with lower masses and line luminosities than our target, allowing us to estimate the value for our star by interpolation rather than extrapolation. We used these accretion luminosities to estimate the mass accretion rate onto the star, using the stellar radius and mass from \citet{2018MNRAS.476.3290M}, and assuming that the inner radius of the accretion disk corresponds to the blackbody radius of the hot disk identified in \citet{2018MNRAS.476.3290M}. The results are listed in Table \ref{table:hydrogen}.
%{\bf Why does this say Table 7.2?  Latex is being weird. -MJK} 
Our mass accretion rates fall $\sim 0.8-2$ dex lower than those found by \citet{2018MNRAS.476.3290M} using the $v_{10}[\mathrm{H\alpha}]-\dot{M}_{\mathrm{acc}}$ relation from \citet{2004A&A...424..603N}. This is expected, given that the $v_{10}[\mathrm{H\alpha}]-\dot{M}_{\mathrm{acc}}$ relation was calibrated for younger stars and was unadjusted for age.

While quantitatively the accretion luminosities do not significantly vary from observation to observation, the accretion signature lines show striking differences in each spectrum, as shown in Figure \ref{fig:Brgamma}. Most notable is the complete disappearance of Br $\gamma$ in the observation of 2018 March 01. While Br $\gamma$ is a shallow line, and thus would be expected to disappear before Pa $\beta$ does, this highlights the need for monitoring in multiple bandpasses at once. The lack of variation in accretion luminosity is due to the conversion from line fluxes to accretion fluxes, which is dominated by uncertainty in the conversion itself (rather than observational uncertainty in the line flux). 

\begin{figure*}
\begin{centering}
\plottwo{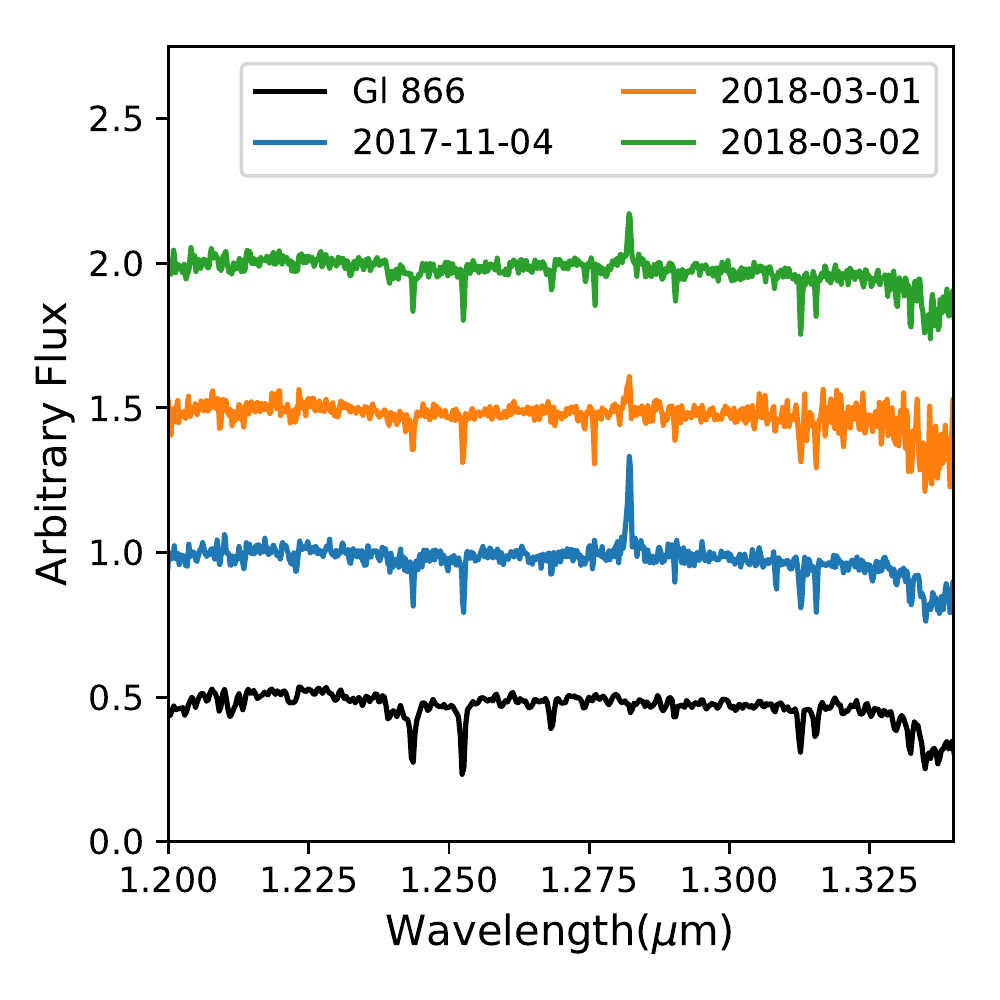}{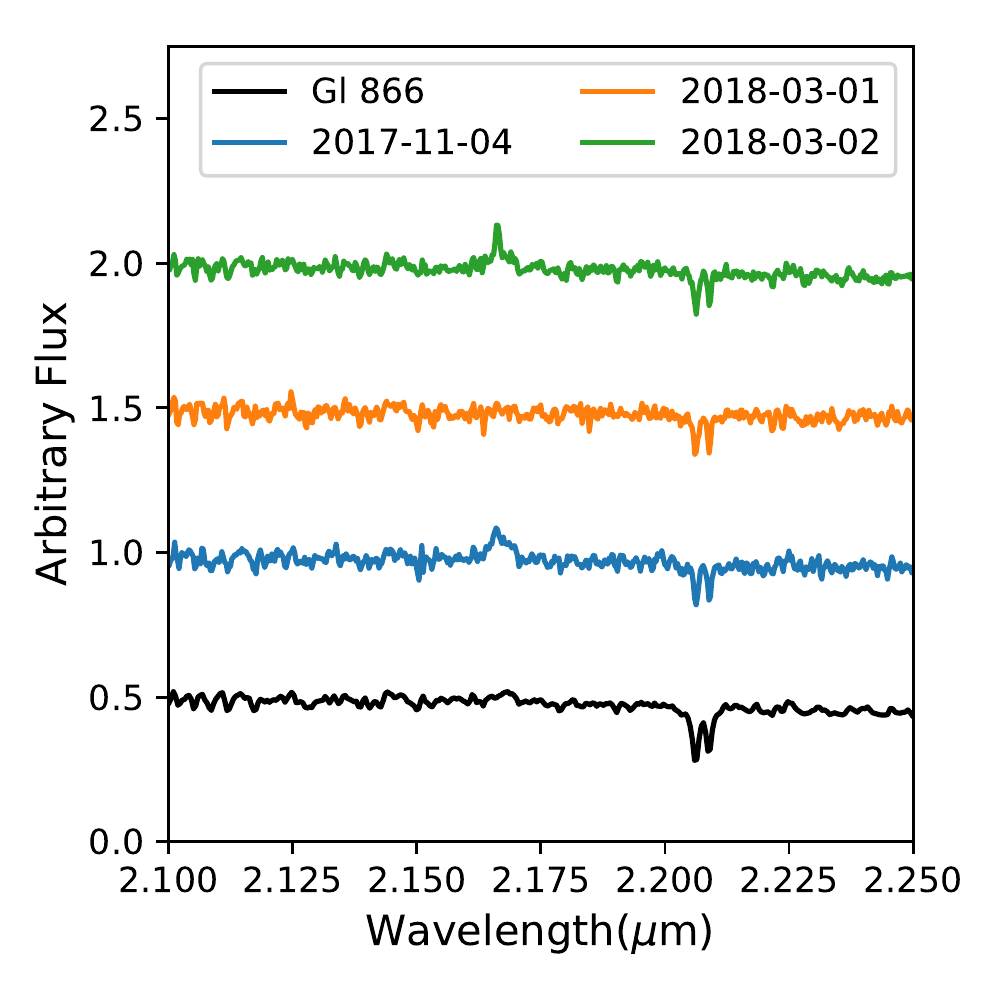}
\caption{\textit{Left}: The Pa$\beta$ region of J0808. \textit{Right}: The Br$\gamma$ region of J0808. Spectra are normalized to the continuum around the line, and separated by an arbitrary offset. The 2017-11-04 spectrum shows a clear peak in Pa$\beta$ and broad emission above the expected underlying continuum (as depicted by Gl 866) in Br$\gamma$. The 2018-03-01 spectrum shows no enhancement whatsoever in Br$\gamma$, and significantly less emission at Pa$\beta$ than in the 2017-11-04 spectrum. The 2018-03-02 spectrum exhibits obvious line profiles in both lines.
\label{fig:Brgamma}}
\end{centering}
\end{figure*}

\section{Characteristics and Formation Mechanisms of Peter Pan Disks}
\label{sec:mechanisms}

In the preceding sections, we have described two new Peter Pan disk systems, and further characterized known Peter Pan disks. Here, we use the accumulated information on these systems to identify the shared characteristics that seem to define a distinct class of objects. We then discuss five potential formation and evolution mechanisms that lead to the observed characteristics of these systems at their ages. We also consider the implications of the findings of \citet{2019ApJ...872...92F}, indicating a lack of CO gas in the system. In the final subsection, we discuss various additional observations that could potentially be used to test these scenarios.

\subsection{What is a Peter Pan Disk?}

There are now eight systems with literature identifications in the Great Austral Young Association (GAYA) that have been claimed as accretion disk candidates. However, there are disagreements in the literature about these targets and their behaviors. We therefore take this opportunity to note the particular characteristics which make these Peter Pan disks a unique population, and summarize each object. 

\begin{figure*}
\plotone{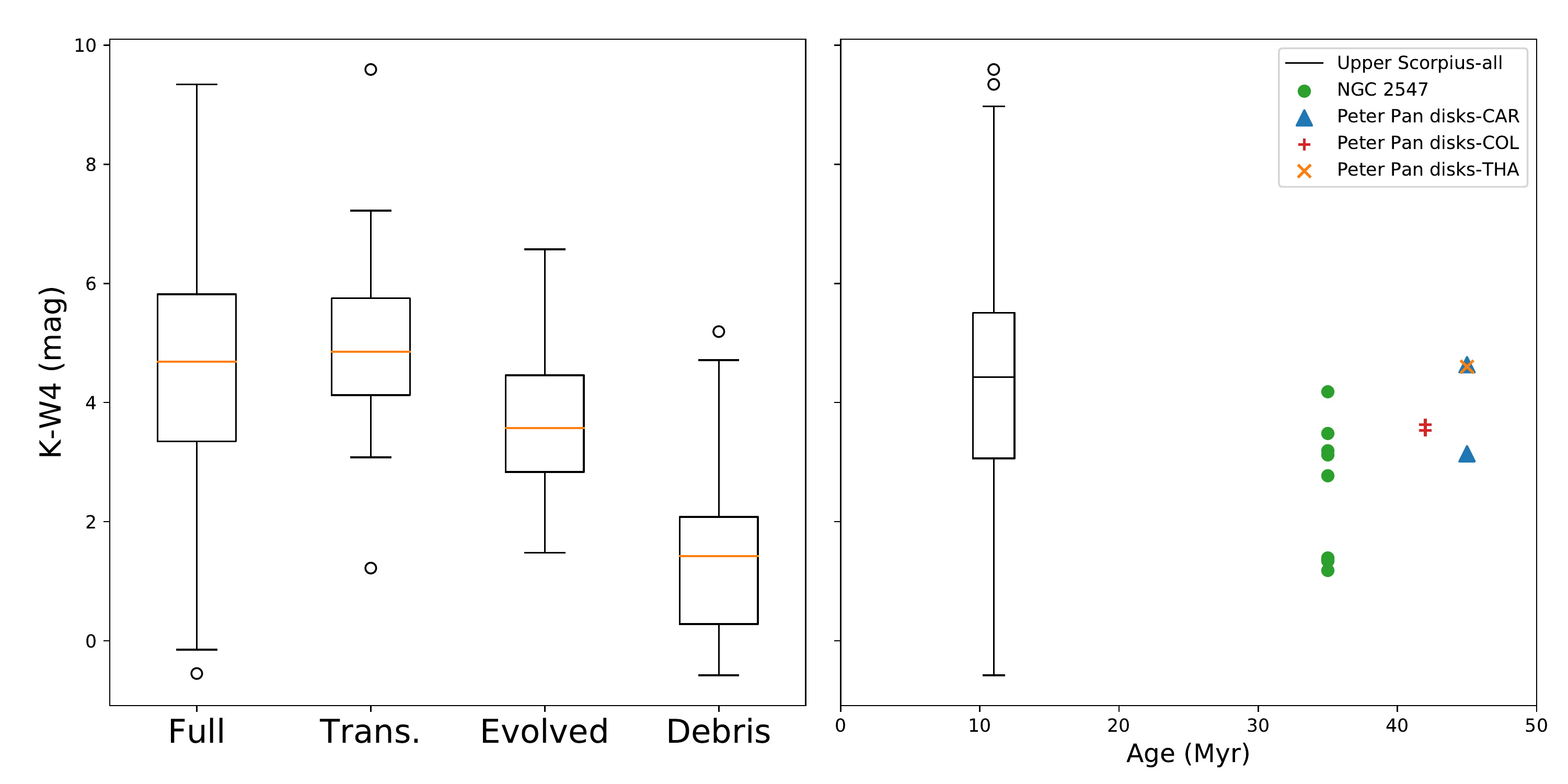}
\caption{\textit{Left}: The range of [K]-[W4] excess for the different classes of M-star disks in Upper Scorpius from \citet{2018AJ....156...75E} shows that debris disks tend to have lower levels of excess than primordial disks even at 10 Myr. Horizontal peach lines represent the median of each set, while boxes indicate the interquartile range (IQR). Open circles represent excess measurements more than 1.5 IQR distant from the nearest quartile. \textit{Right}: The [K]-[W4] colors of Upper Scorpius M-star disks, NGC 2254 M-dwarf disks from \citet{2008ApJ...687.1107F}, and the Peter Pan disk candidates in Table \ref{table:all_disks} show that Peter Pan disks have observed excesses closer to Upper Scorpius primordial disks than to debris disks.}
\label{fig:excess_vs_age}
\end{figure*}

Figure \ref{fig:excess_vs_age} shows the four measured excesses for these systems at $\sim$22 $\mu$m in comparison to the M-type disks identified in the $\sim$10 Myr Upper Scorpius association by \citet{2018AJ....156...71C}, and the sample of M dwarf disks categorized as debris disks by \citet{2008ApJ...687.1107F} in the $\sim$ 35-Myr NGC 2254. This shows that the excesses seen in the Peter Pan disks correspond more closely to ``full,'' ``evolved,'', or ``transitional'' disks, rather than debris disks, following the classification scheme adopted by \citet{2012ApJ...758...31L} and \citet{2014ApJ...784..126E}.\footnote{We note that in this classification scheme, a ``transitional'' disk is a disk with an inner hole or gap, while an ``evolved'' disk is optically thin, but does not have a hole \citep{2012ApJ...758...31L}.}

Thus far, the common features of ``Peter Pan'' disks are:

\begin{enumerate}
    \item a star or brown dwarf of M or later spectral type, %While thus far the earliest spectral type for a Peter Pan disk candidate is M4, we adopt this more conservative boundary to allow for future discoveries. %{\bf why don't we just define it as M or later (since it's much more work to figure out if you have a , and point out that all the examples are mid-Ms? -MJK}
    \item at age $\gtrsim 20$ Myr, based on moving group membership,  %{\bf Looking at Figure 16 OK I do see a gap between 12 and 30 Myr.   But I feel like maybe this gap is there because of a lack of moving groups at those middle ages.  Maybe instead of "20 Myr" you should just say "older than Sco-Cen". -MJK}
    
    %Thus far, there is not enough information about disks in the $20-40$-Myr age range to mark a conclusive lower bound for Peter Pan disks. We thus adopt an approximate boundary based on the $e$-folding timescale for primordial disk fraction in K stars in Sco-Cen \citep{2016MNRAS.461..794P}.
    %{\bf What do you mean "chosen "in part"?  This is the place to state the WHOLE reason for your choice, not "part". -MJK}
    %in a $\sim 45$ Myr moving group %{\bf Why 35 Myr exactly?  Especially given that you're using a 45 Mya moving group?  Couldn't a 40 Myr old moving group have Peter Pan Disks? What if someone discovers a 34 year old moving group? -MJK }
    \item that exhibits substantial mid-IR excess such that the observed color $[K] - [W4] > 2$, chosen based on the observed excess differences between debris and primordial disks in Upper Scorpius \citep[Figure \ref{fig:excess_vs_age}]{2018AJ....156...75E}, %$L_{\mathrm{ir}}/L_{\star} > 0.001$, a common estimate for the boundary between debris and primordial disks \citep{2011ARA&A..49...67W}, {\bf Again, you could refer to Figure 16 and pull your excess criterion from this figure/Sco-Cen.  -MJK} %{\bf It's not "the boundary".  There is no such clear cut boundary.  $L_{\mathrm{ir}}/L_{\star} > 0.001$ is just a common rule of thumb.} 
    %This limit is chosen to firmly establish that these systems are not ``run-of-the-mill'' debris disks.
    %{\bf Hold on--if you invoke the word "extreme" it suggests you're talking about debris disks.  But earlier you said that the disks were bright enough to be primordial disks.  So I recommend you delete this sentence about "extreme" excess. -MJK}
    %at both 12 and 22 microns   {\bf It doesn't make sense to say L{ir}/L{star} AT a given wavelength.  The fractional infrared luminosity (L{ir}/L{star}) is the ratio of integrated infrared excess of the disk to the bolometric luminosity of the star.   If it's at a given wavelength, it's the flux of the disk over the flux of the star, or simply the excess at that wavelength. -MJK}
    \item and has an accretion signature indicative of the presence of warm gas. %though measurements of accretion rate itself are typically dominated by systematic uncertainties \citep{2016ARA&A..54..135H}. %{\bf I think you need to say something about how sensitive such observations are. I.e. what's the lowest measureable Mdot? -MJK} %This requirement differentiates the systems from extreme debris disks \citep[e.g.][]{2005Natur.436..363S,2015ApJ...805...77M}, which thus far are not known to be accreting.
\end{enumerate}

%{\bf This discussion/definition doesn't yet benefit from the box and whisker figure and the accretion rate Figure.   (Figs 16, 18).  I thought you were going to use those to justify criteria 2, 3, and 4 above! I think this "definition" needs to come AFTER those figures. -MJK}

%{\it You should really explain why you have chosen each of these cutoffs.  Probably the best way to do this would be to show a plot of L{ir}/L{star} vs age for M dwarfs, and show that the Peter Pan objects form a distinct cluster, and plot your cutoffs as dotted lines on top. BTW I reordered the paragraphs below about each of the candidate systems and added some bold text to the beginning of each one as a title that it meant to stay. -MJK}

%Thus far, no examples have been identified with a spectral type earlier than M4. We conservatively set our boundary for spectral type as M0 or later to allow for the possibility of future detections.

\textit{J0808, the prototypical example.} J0808 is the prototypical Peter Pan disk. Its kinematics and spectroscopic age indicators place it at an age $\sim 45$ Myr, it has a $L_{\mathrm{ir}}/L_{\star} \sim 0.15$ \citep{2018MNRAS.476.3290M}, and its optical and near-IR spectra show clear evidence of accretion \citep[Section \ref{sec:spectra}]{2018MNRAS.476.3290M}. 

\textit{J0501, another example.} Similar to J0808, J0501 shows clear evidence of age $\sim 42$ Myr, $L_{\mathrm{ir}}/L_{\star} \sim 0.02$, and evidence of accretion from broad H$\alpha$ emission.

{\it J0446AB, a Peter Pan pair.} While both J0446A and J0446B are high-likelihood members of Columba, the H$\alpha$ measurements straddle the boundaries separating an accretion origin from an activity origin defined in the literature \citep{2003ApJ...582.1109W,2003ApJ...592..282J}. We interpret the H$\alpha$ measurements as indicators of accretion and thus identify both targets as Peter Pan disks, but note that measurement of accretion signatures that suffer less from origin confusion (e.g. spectroscopy that resolves the wings of the H$\alpha$ line, near-IR hydrogen emission lines, UV/X-ray observations) is necessary.  

\textit{J0949AB, another likely Peter Pan pair.} J0949A is clearly accreting, based on the presence of He I 6678\AA\ emission and the strength and breadth of the H$\alpha$ emission. While not as clear-cut, J0949B also shows evidence of accretion, based on the breadth of the H$\alpha$ emission.
We thus consider both of these stars Peter Pan disk candidates. As with J0446, additional measures of accretion (UV/X-ray excess, near-IR hydrogen emission) would be beneficial for eliminating confusing with activity-driven H$\alpha$ emission.

\textit{J0226, a brown dwarf Peter Pan candidate.} \citet{2016ApJ...832...50B} also present 2MASS J$02265658-5327032$ (hereafter J0226) as an L0 member of the Tucana-Horologium association (THA), with an age of $45 \pm 4$ Myr. Using updated kinematics for the object from \textit{Gaia} DR2 (including a trigonometric parallax) and the radial velocity from \citet{2016ApJ...832...50B}, J0226 has a 98.8\% likelihood of membership in THA from BANYAN $\Sigma$. The object also exhibits Pa $\beta$ emission indicative of ongoing accretion, though a quantitative measurement of the accretion rate has not yet been made. This object extends the realm of Peter Pan disk down to a much lower mass.

\textit{J0041, a candidate we discard.} While \citet{2014ApJ...783..121G} identified J0041 as a likely member of Tucana-Horologium, \citet{2017AJ....154...69S} identify it as a likely member of the Beta Pictoris moving group. Furthermore, the system does not exhibit any excess with WISE. We thus discard it from consideration as a Peter Pan disk.

\textit{J0949 requires more kinematic information.} Despite an angular separation of $1.49''$ and a physical distance of $\sim1.10 \pm 0.47$ pc (Table \ref{table:new_disks}), the proper motions of the two objects in J0949 significantly differ. J0949B moves $3.039 \pm 0.167$ mas/yr faster than J0949A in right ascension, and $4.98 \pm 0.18$ mas/yr slower in declination. Thus, while J0949A has a membership likelihood in Carina of $>99\%$, J0949B is near-equally likely to be a member of Carina, Lower Centaurus-Crux, or the field. The observed radial velocities make Carina more likely (per BANYAN $\Sigma$), but not overwhelmingly so. One possible explanation is that the observed proper motion difference is due to orbital motion of a bound binary system. It is not possible with the current data to determine if the system is bound, given the large relative uncertainties of the system; the physical separation is only a $2-\sigma$ measurement, while the space velocity relative to the center of mass of the system has uncertainty $>100\%$. Further observations to reduce these kinematic uncertainties (e.g. high-precision RV measurements, improved precision on parallax) will be necessary to confirm this possibility of orbital motion. 

\textit{Further Potential Peter Pan Disks.} Several of the objects in NGC 2254 exhibit $\sim$22 $\mu$m excesses similar to the Peter Pan disks presented here, despite their classification as debris by \citet{2008ApJ...687.1107F}. However, we have not found any published optical or infrared spectra of the NGC 2254 objects with sufficient resolution to measure accretion-caused hydrogen emission from these stars; better spectroscopy of these systems is warranted.

\subsection{Characteristics of Known Peter Pan Disks}
We list the observed characteristics of the seven Peter Pan disk candidates in Table \ref{table:all_disks}, including updated astrometry and moving group membership information for J0808 from \textit{Gaia} DR2. Parallax and proper motion measurements for J0501 were not included in \textit{Gaia} DR2, so we use the characteristics listed in \citet{2016ApJ...832...50B}. In addition to the astrometry, we list accretion rate measurements from both optical and near-IR spectroscopy, WISE colors, and best-fit SED model characteristics of the systems.

\begin{rotate}
\begin{deluxetable*}{lcccccccc}%[tp]
\tabletypesize{\footnotesize}
\tablecaption{Collected Properties of Peter Pan Disks \label{table:all_disks}}
\tablewidth{0pt}
\tablehead{\colhead{Designation} & \colhead{J0808} & \colhead{J0501} & \colhead{J0446A} & \colhead{J0446B} & \colhead{J0949A} & \colhead{J0949B} & \colhead{J0226} & \colhead{Ref.}}
\startdata
R.A. (h:m:s)\tablenotemark{a} & 08:08:22.182 & 05:01:00.889 & 04:46:34.105 & 04:46:34.249 & 09:49:00.753 & 09:49:00.441 & 02:26:56.758 & 1\\
Dec. (d:m:s)\tablenotemark{a} & −64:43:57.26 & -43:37:09.96 & -26:27:56.84 & -26:27:55.57 & -71:38:02.95 & -71:38:03.16 & -53:27:03.46 & 1\\
Distance (pc) & $101.4 \pm 0.6$ & $(47.8_{-8.4}^{+7.2})$\tablenotemark{b} & $82.6 \pm 0.4$ & $82.2 \pm 0.4$ & $79.2 \pm 0.5$ & $78.1 \pm 0.3$ & $46.5 \pm 1.1$ & 1,2 \\
$\mu_{\alpha} \cos \delta$ (mas yr$^{-1}$) & $-11.539 \pm 0.118$ & \nodata\tablenotemark{c} & $33.351 \pm 0.084$ & $33.534 \pm 0.080$ & $-36.096 \pm 0.135$ & $-39.135 \pm 0.099$ & $92.486 \pm 0.773$ & 1,2 \\
$\mu_{\delta}$ (mas yr$^{-1}$) & $25.615 \pm 0.100$ & \nodata\tablenotemark{c} & $-5.459 \pm 0.118$ & $-3.629 \pm 0.112$ & $28.565 \pm 0.131$ & $23.582 \pm 0.117$ & $-21.347 \pm 1.137$ & 1,2 \\
RV (km s$^{-1}$) & $22.7 \pm 0.5$ & $19.6 \pm 0.5$ & $26.7 \pm 16.8$ & $29.8 \pm 16.8$ & $22.4 \pm 16.7$ & $20.5 \pm 16.8$ & $4.7 \pm 3.3$ & 3,2,4 \\
Spectral Type & M5 & M4.5 & M6 & M6 & M4 & M5 & L0 $\delta$ & 3,2,4 \\
Association & CAR & COL & COL & COL & CAR & CAR & THA & 5,2,4 \\
Age (Myr) & $45_{-7}^{+11}$ & $42_{-4}^{+6}$ & $42_{-4}^{+6}$ & $42_{-4}^{+6}$ & $45_{-7}^{+11}$ & $45_{-7}^{+11}$ & $45 \pm 4$ & 6 \\
$\log_{10}(\dot{M}_{\mathrm{acc,H\alpha}}$ $(M_{\odot}$ $ \mathrm{yr^{-1}}))$ & $(-10)-(-9.5)$ & $-10.80_{-0.05}^{+0.07}$ & $-10.9 \pm 0.4$ & $-10.6 \pm 0.4$ & $-9.3 \pm 0.4$ & $-9.9 \pm 0.4$ & \nodata & 3,2,4 \\
$\log_{10}(\dot{M}_{\mathrm{acc,IR}}$ $(M_{\odot}$ $ \mathrm{yr^{-1}}))$ & (-12)-(-10.8) & \nodata & \nodata & \nodata & \nodata & \nodata & \nodata & 4 \\
$[\mathrm{W1}]-[\mathrm{W3}]$ (mag) & $2.445 \pm 0.036$ & $1.256 \pm 0.039$ & \multicolumn{2}{c}{$1.146 \pm 0.034$} & \multicolumn{2}{c}{$1.468 \pm 0.027$} & $1.7 \pm 0.2$ & 4 \\
$[\mathrm{W1}]-[\mathrm{W4}]$ (mag) & $4.312 \pm 0.087$ & $3.42 \pm 0.08$ & \multicolumn{2}{c}{$3.093 \pm 0.078$} & \multicolumn{2}{c}{$2.888 \pm 0.043$} & $4.6 \pm 0.3$ & 5,2,4 \\
$T_{\mathrm{eff}}$ (K) & $3050 \pm 100$ & 3125 & $\sim3000$\tablenotemark{d} & $\sim3000$\tablenotemark{d} & $\sim 3200$\tablenotemark{d} & $\sim3050$\tablenotemark{d} & 2260\tablenotemark{e} & 3,2 \\
$T_{\mathrm{disk,hot}}$ (K) & $1071 \pm 103$ & \nodata & \nodata & \nodata & \nodata & \nodata & \nodata & 3 \\
$L_{\mathrm{disk,hot}}/L_{\star}$ & $0.054 \pm 0.018$[0...0.11]\tablenotemark{f} & \nodata & \nodata & \nodata & \nodata & \nodata & \nodata & 3 \\
$T_{\mathrm{disk,warm}}$ (K) & $237 \pm 11$ & $170 \pm 10$ &  \multicolumn{2}{c}{$299^{+11}_{-9}$}  & \multicolumn{2}{c}{$279^{+7}_{-6}$} & $135 \pm 20$ & 3,2,4 \\
$L_{\mathrm{disk,warm}}/L_{\star}$ & $0.070 \pm 0.015$ & $0.021_{-0.001}^{+0.002}$ & \multicolumn{2}{c}{$0.0262 ^{+0.0006}_{-0.0008}$\tablenotemark{g}} & \multicolumn{2}{c}{$0.0178 \pm 0.0003$\tablenotemark{g}} & $0.15^{+0.06}_{-0.04}$ & 3,2,4 \\
$T_{\mathrm{disk,cold}}$ (K) & $>20$& \nodata & \nodata & \nodata & \nodata & \nodata & \nodata & 7 \\
\enddata
\tablenotetext{a}{Right ascension and declination are the stellar positions at epoch J2015.5, the \textit{Gaia} reference epoch.}
\tablenotetext{b}{Statistical distance from \citet{2016ApJ...832...50B}.}
\tablenotetext{c}{No trigonometric proper motion data available from \textit{Gaia}.}
\tablenotetext{d}{$T_{\mathrm{eff}}$ adopted from temperature of best fit BT-Settl stellar model.}
\tablenotetext{e}{$T_{\mathrm{eff}}$ from $T_{\mathrm{SpType}}$ presented in \citet{2016ApJ...832...50B}.}
\tablenotetext{f}{Approximate range from variability in W1 and W2 photometry \citep[see][]{2018MNRAS.476.3290M}.}
\tablenotetext{g}{$L_{\mathrm{ir}}/L_{\star}$ calculated using combined flux from both stars.}

\tablerefs{(1) \citet{2018AandA...616A...1G} (2) \citet{2016ApJ...832...50B} (3) \citet{2018MNRAS.476.3290M} (4) This paper (5) \citet{2016ApJ...830L..28S} (6) \citet{2015MNRAS.454..593B} (7) \citet{2019ApJ...872...92F}}

\end{deluxetable*}
\end{rotate}

\clearpage

\subsection{Origins of Peter Pan Disks}
Below, we discuss six possible origins for Peter Pan disks.  The most straightforward scenario is that these systems are long-lived primordial disks, and that our understanding of M dwarf disk evolution is still emerging. However, we also consider the possibility that some M dwarfs host secondary disks that form long after the primordial disk dissipates, analogous to disks around white dwarfs or other evolved stars.

\subsubsection{Slow Dissipation of Primordial Disks about Low-Mass Stars}

The simplest explanation for the existence of Peter Pan disks is that, in general, M dwarfs simply dissipate their primordial disks less rapidly than higher mass stars. Perhaps they are part of the trend observed in stars more massive than M dwarfs: disks around stars of masses $> 1.2 M_{\odot}$ dissipate in approximately half the time as stars with masses below this \citep[e.g.][]{2006ApJ...651L..49C}, likely due to their higher accretion rates \citep[e.g.][]{2005AJ....129..935C} and higher levels of X-ray photoevaporation \citep[e.g.][]{2012MNRAS.422.1880O}. The low accretion rates identified in this work compared to other primordial disks around stars with similar masses (albeit with larger radii) would correspond to this difference, but on the low-mass end. 

The observed primordial disk fraction of mid-late M dwarfs (spectral types later than M3.5) in the $\sim 10$ Myr Upper Scorpius association from \citet{2018AJ....156...71C} corresponds to an $e$-folding timescale of $\sim 9.3$ Myr. 
Extrapolating to the ages of the groups considered here, we would expect an occurrence rate of Peter Pan disks of $\sim 1\%$ for Columba, and $\sim 0.8\%$ for Carina and Tucana-Horologium.   
It is likely that such systems have not been detected until now due to the lack of precise astrometry for late-type objects in the pre-\textit{Gaia} era---one would expect that the low-mass ends of moving groups will be better characterized now that this data is available. Indeed, such progress is underway \citep[e.g.][]{2018ApJ...862..138G}, but additional observations (radial velocity measurements and independent age indicators) are necessary to confirm membership lists for these associations.

\begin{figure*}
\plottwo{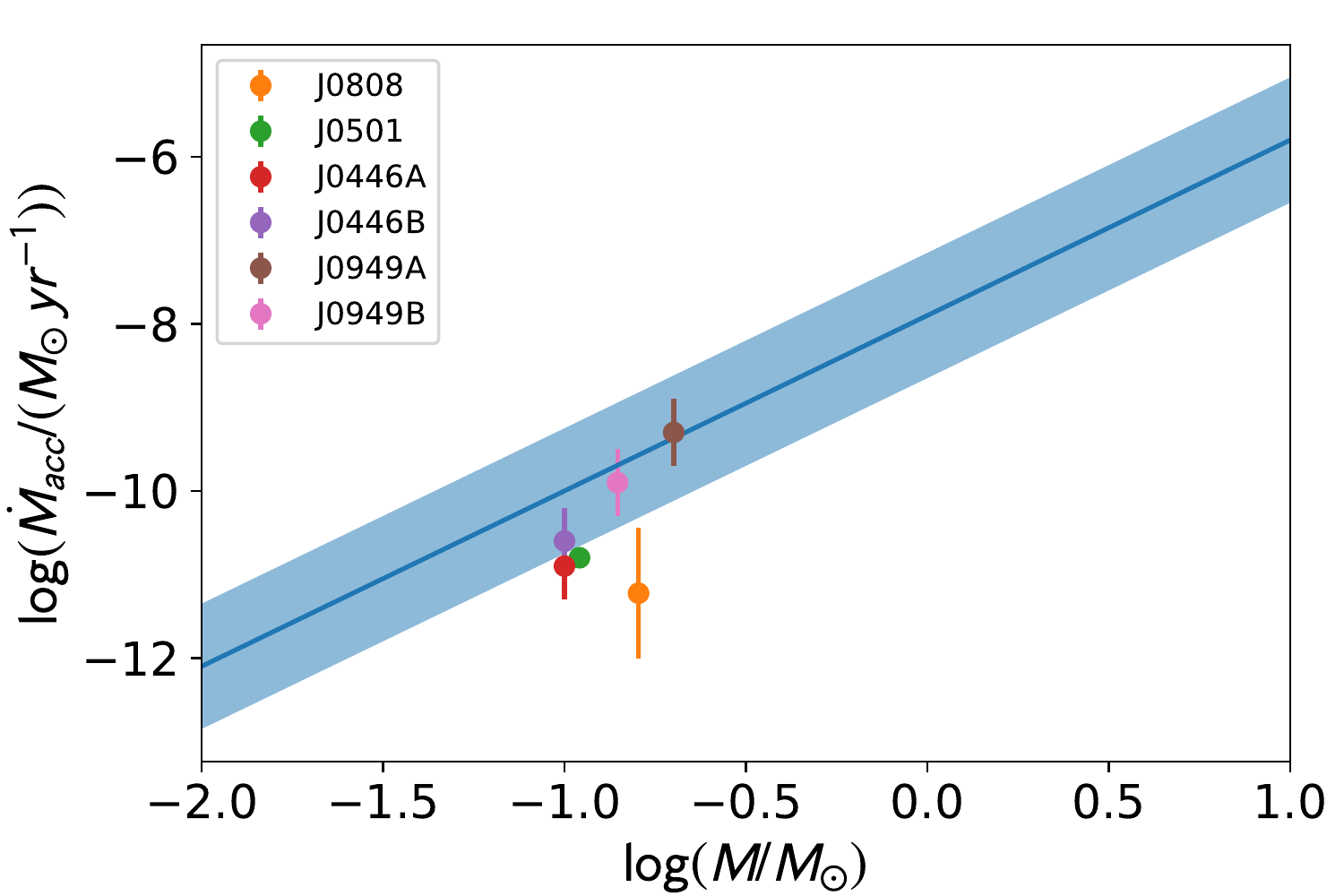}{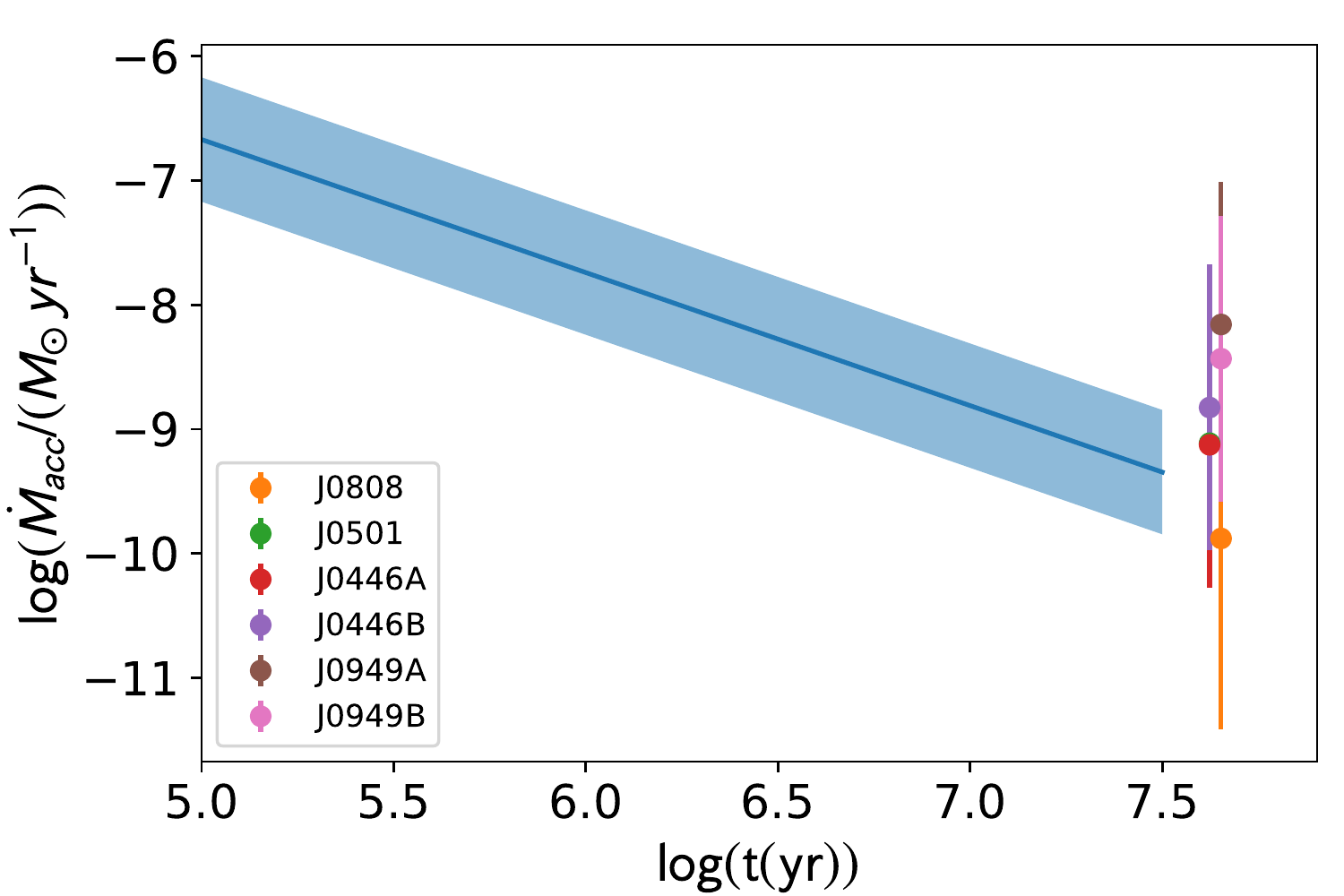}
\caption{\textit{Left}: Accretion rate as a function of mass for the Peter Pan disks with measured accretion rates in this work, compared to the accretion rate-mass relation from \citet{2016ARA&A..54..135H}. Accretion in Peter Pan disks is consistent with typical pre-main sequence accretion. \textit{Right}: Accretion rate as a function of age for Peter Pan disks, scaled to $0.7\mathrm{M_{\odot}}$, compared to the relation from \citet{2016ARA&A..54..135H}. Peter Pan disks appear to extend this relation to an older age.}
\label{fig:accretion_rates_vs_hartmann}
\end{figure*}

Figure \ref{fig:accretion_rates_vs_hartmann} shows the observed accretion rates for our sample of Peter Pan disks against the accretion rate-mass and accretion rate-age relations derived by \citet{2016ARA&A..54..135H}. \citet{2016ARA&A..54..135H} present a linear relationship between $\log(\dot{M}_\mathrm{acc})$ and $\log(M_{\star})$ of

\begin{equation}
    \log(\dot{M}_\mathrm{acc}) = - 7.9 + 2.1 \times \log(M_{\star})
\end{equation}

\noindent with scatter $\sim 0.75$ dex. The accretion rates measured for the Peter Pan disks are consistent with this relationship, assuming a stellar mass for the objects corresponding to the main-sequence mass of a star of that spectral type for the new objects. Following \citet{2016ARA&A..54..135H}, we then scaled the accretion rates of the Peter Pan disks to a stellar mass of $0.7M_{\odot}$:

\begin{equation}
    \log(\dot{M}_\mathrm{acc,scale}) = \log(\dot{M}_\mathrm{acc}) + 2.1 \log(0.7M_{\odot}/M_{\star})
\end{equation}

\noindent
These scaled accretion rates at $\sim 45$ Myr are consistent with the accretion-rate-time relationship from \citet{2016ARA&A..54..135H}, which was calibrated on stars of ages $<30$ Myr. We note the caveat from \citet{2016ARA&A..54..135H} that the sample may be biased due to correlated uncertainties in age and accretion rate. We also note that five objects with masses $>0.3 M_{\odot}$ in \citet{2016ARA&A..54..135H} have isochronal ages of $20-30$ Myr despite apparent membership in much younger associations. Given the apparent discrepancy in age depending on group membership, we thus continue to adopt an approximate lower age limit for Peter Pan disks of $\sim 20$ Myr.

A long-lived primordial disk would be expected to exhibit cold gas, as well as dust. ALMA observations of J0808 \citep{2019ApJ...872...92F} do not show any evidence of CO gas, the most common tracer. However, \citet{2019ApJ...872...92F} demonstrate that CO gas may photodissociate around M dwarfs on timescales less than the ages of this system, suggesting that this non-detection has limited impact on our understanding of the origin of this disk.  
Additional submillimeter observations would be useful to  characterize the presence of non-CO gas (e.g. H$_{2}$O) in the system and in other Peter Pan disk systems. These observations could also trace the presence of [CI], which has been shown to shield CO from photodissociation \citep{2019MNRAS.489.3670K}; the presence of [CI] could indicate that a dissipation method other than photodissociation is at play.

Viewing Peter Pan disks as long lived primordial disks raises some nagging questions: why have disks around these particular stars survived so long?
Were these systems shielded from high-energy radiation external to the system \citep[e.g.][]{2013ApJ...774....9A}? Did their host stars produce atypically little X-ray and UV radiation?   Better characterizing the UV activity and chemistry of Peter Pan disk host stars could provide insights into why the disks around these particular stars seem to have lingered around for so long.

\subsubsection{Hybrid Disks or Extreme Debris Disks, Arising from Recent Collisions}

Recent observations of J0808 with ALMA \citep{2019ApJ...872...92F} show optically-thin dust at 1.3 mm and no evidence of CO.  This combination led \citet{2019ApJ...872...92F} to suggest that the system is not a conventional primordial disk, but a gas-free debris disk. They suggested that dust grains released in the outer disk could migrate inwards via Poynting-Robertson (PR) drag, potentially serving as a source of grains for the strong infrared excess. If this were true, J0808, with its fractional infrared luminosity of $>0.05$ would be a very "extreme" debris disk. Indeed \citet{2019ApJ...872...92F} estimate that if the dust mass estimated from the 1.3-mm emission were to migrate inwards to this inner radius it would be highly optically thick (optical depth $\tau \sim 30000$).

However, we find this PR transport scenario implausible, because of the fundamental limit that grain-grain collisions impose on the optical depth in a transport-dominated debris disk \citep{2010AJ....140.1007K}. Grains transported by PR drag or any kind of drag can not produce a grain pileup with a face-on optical depth greater than the critical value of $\tau \sim v_K/c$, where $v_K$ is the local Keplerian speed \citep{2010AJ....140.1007K}.  For dust orbiting 0.01 au from an M dwarf, this limit is roughly $\tau \sim 3 \times 10^{-4}$.  The fractional infrared luminosity may differ from the face-on optical depth because of geometrical effects, but these are not likely to provide more than an order of magnitude of flexibility. PR drag driven by a stellar wind or scattered radiation could operate in these systems. However, to effectively transport dust in disks with an optical depth greater than $v_{\mathit{Kepler}}/c$, transport forces would need to operate on timescales shorter than the collision time, which these secondary forces do not.

A more likely scenario, which \citet{2019ApJ...872...92F} also discuss, would be that the disk around J0808 (and perhaps other Peter Pan disks) arises from a recent collision among protoplanets, or a new collisional cascade of asteroid analogs. The dipper-like variability we observed in J0808 with CTIO and TESS may support this scenario.

However, observations of accretion signatures in spectra of J0808 by both \citet{2018MNRAS.476.3290M} and by us imply an accretion rate of  $0.3-5 M_{\oplus}$ of hydrogen per Myr. Debris disks are generally considered to be mostly free of hydrogen gas, although \citet{2017A&A...599A..75W} detected hydrogen in the $\beta$~Pictoris debris disk.  If J0808 is a debris disk, it is most similar to the family of gas-rich debris disks, or perhaps the subset of these that is sometimes referred to as "hybrid disks" \citep{2017A&A...600A..62P,2017ApJ...849..123M}, which may retain some primordial gas.

\subsubsection{Star-Star and Star-Disk Interactions}

Could secondary Peter Pan disks arise from stellar collisions or close encounters?  During a close encounter, the M dwarf could acquire the disk of the other star \citep[e.g.][]{2005ApJ...629..526P}.  Alternatively an M dwarf could collide with a lower-mass M dwarf or brown dwarf, forming a disk from the high-angular momentum material \citep[e.g.][]{1995ApJ...447L.121L}. However, such stellar encounters rarely occur outside of a tightly packed cluster (young moving groups are not nearly dense enough), and direct stellar collisions tend to produce a younger-looking, more massive stars (i.e. not M dwarfs).

\subsubsection{Tidal disruption of a giant planet}

Could secondary Peter Pan disks be fed by tidal disruption of gas giant planets?  Perhaps, but the geometry of the resulting disk seems be a poor match for the disks we observe. The Roche radius for J0808 and a planet with Neptune's density and radius is $2 \times 10^{-4}$ au. This radius is well inside the $5.6 \times 10^{-3}$ au blackbody radius of J0808's inner disk \citep[as identified by][]{2018MNRAS.476.3290M}, and the other disks in Table \ref{table:all_disks} are not significantly closer to their respective stars.

\subsubsection{A disk due to a mass transfer binary}

Could secondary Peter Pan disks arise from mass transfer in a binary system?
This scenario requires a ``primary'' star massive enough to evolve onto and off the main sequence in $<45$ Myr while low enough mass to evolve into a white dwarf, rather than producing a core-collapse supernova. Given that the main-sequence lifetime for a B1V star, too massive to produce a white dwarf, is $\sim 45$ Myr, this scenario can not explain the disks we observer in Carina, Columba, and Tuc-Hor. However, it could potentially generate Peter Pan disks around older M dwarfs in older groups, if those were ever identified.

\subsection{More Observations Needed}

Our understanding of the parameters, membership and the origin of this apparent new category of disks would benefit from further observations, in addition to the sub-mm and UV follow-up we suggested above.

\begin{itemize}

\item \textit{Radio/sub-mm observations}: Primordial disks are expected to have substantial reservoirs of cold material. Observations at sub-mm wavelengths in multiple bands would identify whether such a reservoir exists in these systems. Such observations would also enable estimation of the system's disk mass and gas content, critical for understanding the evolutionary state of the system. If no gas is detected in the system, it could indicate a formation scenario closer to that described by \citet{2019ApJ...872...92F}. Multiple bands are necessary to constrain the temperature of the emission, and to probe multiple molecules (e.g. CO and $\mathrm{H_{2}O}$). 

\item \textit{High-energy observations}: 
A key driver of primordial disk dissipation is photoevaporation of the disk's gas content from far- and extreme-ultraviolet (FUV and EUV) radiation \citep[e.g.][]{2011ARA&A..49...67W}. Studies of the low-mass population of the TW Hya association have shown an anti-correlation between X-ray luminosity and disk fraction as a function of spectral type; earlier M dwarfs have higher X-ray luminosities and a lower disk fraction than mid-late M dwarfs \citep{2016AJ....152....3K}. If Peter Pan disks are long-lived primordial disks, this observed anticorrelation with spectral type would provide an explanation for why they have not been detected around earlier spectral types. X-ray observations of the ensemble of Peter Pan disks would enable measurement of the current photoevaporation rate to demonstrate if it is in fact lower than expected, allowing a longer disk life than typically found for primordial disks.

Additionally, high-energy spectroscopy of Peter Pan disks could yield information as to the composition of the accretion material \citep[e.g.][]{2002ApJ...567..434K, 2004A&A...418..687S}. If the accretion material is the gas expected from a classical T Tauri star, that would provide support for the long-lived primordial disk hypothesis. An accretion material composition similar to vaporized exocometary material, alternatively, would support the \citet{2019ApJ...872...92F} theory of an evolved disk, with substantial excesses in the WISE bands due to migration of large grains inward. 

\item \textit{High-resolution spectroscopy} Higher-resolution ($\gtrsim 7000$) spectroscopy of a larger sample of optical Peter Pan disks would allow us to firmly resolve accretion behavior in the spectra, such as line asymmetries in H$\beta$\citep{2016ARA&A..54..135H}, and identification of other narrow emission lines indicative of accretion, e.g. He I $\lambda5786$, [O I] $\lambda 6300$, and He I $\lambda 6678$ \citep{2016ARA&A..54..135H,2018MNRAS.476.3290M}. High-resolution spectra will also enable detection of the narrow Li I line at 6708 \AA\ to further confirm the ages of the stars.

\item \textit{Mid-infrared spectroscopy}: \citet{2019ApJ...872...92F} note that observation of silica features could test the planetary-body collision hypothesis \citep{2009ApJ...701.2019L}. A monitoring campaign of such observations would also enable constraint of the variable excess detected in W1/W2.

\item \textit{Broad-band photometry and light curves}: Near-UV, $U$-, or $u$-band observations of Peter Pan disks would allow us to detect continuum emission from the accretion material \citep{2016ARA&A..54..135H}. Detection of an excess in this range would more clearly indicate accretion, rather than chromospheric activity, as the primary driver of H$\alpha$.

Optical light curves of these systems can provide information as to the activity level of the system (e.g. flares), alignment of the system relative to the line of sight (see Section \ref{sec:tess_J0501}), and disk interactions \citep[e.g. bursting and dipping;][]{2014AJ....147...82C}. As \textit{TESS} continues into its extended mission, its light curves will provide additional constraints on Peter Pan disks as more are observed.

\end{itemize}

\section{Summary}
\label{sec:conclusion}

In this paper, we proposed a set of observational characteristics of a new class of ``Peter Pan'' disk systems: an M-type or later star or brown dwarf with a substantial, warm excess, at a stellar age $>20$ Myr, and exhibiting ongoing accretion. 
Four new Peter Pan disk detections, based on a crossmatch of \textit{Gaia} DR2 with the Disk Detective input catalog and testing objects with the BANYAN $\Sigma$ software, were presented. High-cadence photometric observations of WISEA J0808, the prototypical Peter Pan disk, with the CTIO 0.9m telescope and TESS revealed some flare activity and dipping events, attributable to disk occultations. High-cadence TESS light curves of 2MASS J0501, another Peter Pan disk, shows evidence for a persistent starspot or starspot complex and ongoing flare activity. Using the period from the TESS observations and a projected rotational velocity from the literature \citep{2016ApJ...832...50B}, we found J0501 to have an inclination angle of $\sim 38^{\circ}$. We presented near-infrared spectroscopy of J0808 over six months, showing low-level accretion variable on 24-hour timescales. With these observations in mind, we discussed several potential formation mechanisms for these systems, before concluding that the two most likely mechanisms were long-lived primordial disks, and a recent collisional cascade of protoplanets combined with a long-lived gas component to produce the observed WISE excess. In either case, a unusually long-lived gas component seems to be required to explain the observed accretion onto the star, which raises the question of why gas has persisted around these particular stars.

 %{\bf This needs revising. We do NOT think they are "evolved" dust disks with large grains drawn inward by PR drag. As we say in Section 8.3.2, we find this PR transport scenario implausible, but a hybrid disk scenario is OK. Let's reiterate that here. -MJK}

\citet{2019ApJ...872...92F} noted that J0808 was an oddity, given its substantial excess and ongoing accretion. The identification of four new candidate systems with similar characteristics, in addition to the three previously-known Peter Pan disks, throws this categorization into question. It seems plausible that that the dearth of detections of disks of this sort is instead due to the limits of observation, rather than an actual dearth of such systems. As \textit{Gaia} measures the astrometry of more low-mass stars, more low-mass moving group members will be identified, enabling further identification of such systems.

Despite the approach to disk identification that Disk Detective adopts, investigating every object with potential excess, M dwarf disks remain elusive. However, Peter Pan disks as a class likely would not have been identified without such an approach, as the systems identified thus far would not have been included in a typical WISE disk search. Incorporation of new resources that have become available since the launch of Disk Detective, such as Pan-STARRS \citep{2016arXiv161205560C} and unWISE \citep{2014AJ....147..108L}, may significantly increase the yield of Peter Pan disks in the near future. 

\acknowledgements

We thank the anonymous reviewer for providing comments  that  helped  to  improve  the  content  and  clarity  of  this  paper. The authors thank Anne Boucher (iREX) for graciously providing the spectrum of 2MASS J0501 to use for comparison to our targets. S.M.S. thanks Todd Henry of the SMARTS consortium for his invaluable advice on the optimal usage of the CTIO 0.9m telescope for this project, and Jonathan Gagn\'e for insights interpreting the results of BANYAN $\Sigma$. The authors thank Nathan Kaib for helpful discussions of planetary system dynamics in binary systems, and Scott Kenyon for valuable comments. The authors acknowledge support from grant 14-ADAP14-0161 from the NASA Astrophysics Data Analysis Program and grant 16-XRP16\_2-0127 from the NASA Exoplanets Research Program. M.J.K. acknowledges funding from the NASA Astrobiology Program via the Goddard Center for Astrobiology.

This publication uses data generated via the Zooniverse.org platform, development of which is funded by generous support, including a Global Impact Award from Google, and by a grant from the Alfred P. Sloan Foundation.

Based in part on observations obtained at the Gemini Observatory, acquired through the Gemini Observatory Archive and processed using the Gemini IRAF package, which is operated by the Association of Universities for Research in Astronomy, Inc., under a cooperative agreement with the NSF on behalf of the Gemini partnership: the National Science Foundation (United States), National Research Council (Canada), CONICYT (Chile), Ministerio de Ciencia, Tecnolog\'{i}a e Innovaci\'{o}n Productiva (Argentina), Minist\'{e}rio da Ci\^{e}ncia, Tecnologia e Inova\c{c}\~{a}o (Brazil), and Korea Astronomy and Space Science Institute (Republic of Korea).

Based in part on observations at Cerro Tololo Inter-American Observatory, National Optical Astronomy Observatory (2017A-0259; PI: S. Silverberg; 2017B-0229; PI: S.Silverberg; 2018A-0292; PI: S. Silverberg), which is operated by the Association of Universities for Research in Astronomy (AURA) under a cooperative agreement with the National Science Foundation. 

This publication makes use of data products from the Wide-Field Infrared Survey Explorer, which is a joint project of the University of California, Los Angeles, and the Jet Propulsion Laboratory (JPL)/California Institute of Technology (Caltech), and NEOWISE, which is a project of JPL/Caltech. WISE and NEOWISE are funded by NASA.

2MASS is a joint project of the University of Massachusetts and the Infrared Processing and Analysis Center (IPAC) at Caltech, funded by NASA and the NSF. 

The Digitized Sky Survey was produced at the Space Telescope Science Institute under U.S. Government grant NAG W-2166. The images of these surveys are based on photographic data obtained using the Oschin Schmidt Telescope on Palomar Mountain and the UK Schmidt Telescope. The plates were processed into the present compressed digital form with the permission of these institutions.

The Pan-STARRS1 Surveys (PS1) and the PS1 public science archive have been made possible through contributions by the Institute for Astronomy, the University of Hawaii, the Pan-STARRS Project Office, the Max-Planck Society and its participating institutes, the Max Planck Institute for Astronomy, Heidelberg and the Max Planck Institute for Extraterrestrial Physics, Garching, The Johns Hopkins University, Durham University, the University of Edinburgh, the Queen's University Belfast, the Harvard-Smithsonian Center for Astrophysics, the Las Cumbres Observatory Global Telescope Network Incorporated, the National Central University of Taiwan, the Space Telescope Science Institute, the National Aeronautics and Space Administration under Grant No. NNX08AR22G issued through the Planetary Science Division of the NASA Science Mission Directorate, the National Science Foundation Grant No. AST-1238877, the University of Maryland, Eotvos Lorand University (ELTE), the Los Alamos National Laboratory, and the Gordon and Betty Moore Foundation.

This research has made use of the SIMBAD database, operated at CDS, Strasbourg, France. Some of the data presented in this paper were obtained from the Mikulski Archive for Space Telescopes (MAST). STScI is operated by the Association of Universities for Research in Astronomy, Inc., under NASA contract NAS5-26555. Support for MAST for non-HST data is provided by the NASA Office of Space Science via grant NNX13AC07G and by other grants and offices. This research has made use of the VizieR catalogue access tool, CDS, Strasbourg, France.

IRAF is distributed by the National Optical Astronomy Observatory, which is operated by the Association of Universities for Research in Astronomy (AURA) under a cooperative agreement with the National Science Foundation. PyRAF is a product of the Space Telescope Science Institute, which is operated by AURA for NASA. This research made use of ds9, a tool for data visualization supported by the Chandra X-ray Science Center (CXC) and the High Energy Astrophysics Science Archive Center (HEASARC) with support from the JWST Mission office at the Space Telescope Science Institute for 3D visualization.

\software{IRAF \citep{1993ASPC...52..173T}, PyRAF, AstroPy \citep{2013A&A...558A..33A}, NumPy \citep{van2011numpy}, SciPy \citep{jones_scipy_2001}, Matplotlib \citep{Hunter:2007}, pandas \citep{mckinney}, AstroImageJ \citep{2017AJ....153...77C}, L.A.Cosmic \citep{2001PASP..113.1420V}, Spextool \citep{2003PASP..115..389V,2004PASP..116..362C}, PyVAN \citep{2019arXiv190303240L}}

\facilities{CTIO:2MASS, WISE, Gaia, Gemini:South, CTIO:0.9m, TESS, Blanco:ARCoIRIS}

\bibliographystyle{aasjournal}
\bibliography{apj-jour,references}

%% This command is needed to show the entire author+affilation list when
%% the collaboration and author truncation commands are used.  It has to
%% go at the end of the manuscript.
%\allauthors

%% Include this line if you are using the \added, \replaced, \deleted
%% commands to see a summary list of all changes at the end of the article.
%\listofchanges

\end{document}